\documentclass[preprint,1p,11pt]{IR-Template/ISAS_IR}

\usepackage[english]{babel}
\usepackage[utf8]{inputenc}
\usepackage{tikz}
\usepackage{subcaption}
\usepackage{booktabs}
\usepackage{setspace}
\usepackage{microtype}

\usetikzlibrary{calc,arrows,decorations.pathreplacing}

\usepackage{amsmath}
\usepackage{amssymb}
\usepackage{amsthm}
\usepackage{dsfont}
\usepackage{mathtools}
\usepackage{algorithm}
\usepackage{algorithmic}
\usepackage{xspace}
\usepackage{todonotes}

\newcommand{\sskf}{S\textsuperscript{2}KF\xspace}
\newcommand{\ckf}{fifth-degree CKF\xspace}

\mathtoolsset{showonlyrefs}

\newcommand{\IR}{\mathds{R}}

\newcommand{\IRplus}{\IR_{+}}

\newcommand{\IN}{\mathds{N}}

\renewcommand{\vec}[1]{\underline{#1}}
\newcommand{\vzero}{\vec{0}}

\newcommand{\mat}[1]{\mathbf #1}

\newcommand{\mzero}{\mat{0}}

\newcommand{\mA}{\mat{A}}

\newcommand{\mC}{\mat{C}}

\newcommand{\mI}{\mat{I}}

\newcommand{\mR}{\mat{R}}

\newcommand{\T}{^{\top}}

\newcommand{\dd}{\operatorname{d}\!}

\newcommand{\inv}[1]{\left(#1\right)^{-1}}
 
\DeclareMathOperator{\diag}{diag}

\newcommand{\norm}[1]{{\|#1\|}_2}
\newcommand{\sqNorm}[1]{\norm{#1}^2}

\DeclareMathOperator{\Ei}{Ei}

\newcommand{\mean}[1]{\hat{#1}}

\newcommand{\E}[2][]{{\mathbb{E}}_{#1}[#2]}

\newcommand{\cond}{\,|\,}

\newcommand{\Gauss}{\mathcal{N}}

\newcommand{\JointGauss}[8]{
    \Gauss\left(\Bmat #1 \\ #2 \Emat\hspace{-0.1cm};\hspace{-0.1cm}
           \Bmat #3 \\ #4 \Emat\hspace{-0.1cm},\hspace{-0.1cm}
           \Bmat #5 & \hspace{-0.3cm} #6 \\ #7 & \hspace{-0.3cm} #8 \Emat\right)
}

\newcommand{\gaussExp}[2]{\exp\left(-\frac{1}{2}\frac{#1}{#2}\right)}

\newcommand{\Uniform}{\mathcal{U}}

\newcommand{\LCDN}{F_{\Gauss}}
\newcommand{\LCDDMEven}{F_{\delta}^e}
\newcommand{\LCDDMOdd}{F_{\delta}^o}
\newcommand{\bMax}{b_{\text{max}}}

\newcommand{\vc}{\vec{c}}

\newcommand{\ve}{\vec{e}}
\newcommand{\vm}{\vec{m}}

\newcommand{\vz}{\vec{z}}
\newcommand{\vnu}{\vec{\nu}}

\newcommand{\vck}{\vc_{k}}
\newcommand{\vmk}{\vm_{k}}

\newcommand{\vnuk}{\vnu_{k}}

\newcommand{\evc}{\mean{\vc}}

\newcommand{\vx}{\vec{x}}                   
\newcommand{\vxk}{\vx_{k}}                  
\newcommand{\vxkk}{\vx_{k-1}}               

\newcommand{\evx}{\mean{\vx}}               

\newcommand{\evxkp}{\evx_{k}^{p}}           
\newcommand{\evxke}{\evx_{k}^{e}}           
\newcommand{\evxkke}{\evx_{k-1}^{e}}        

\newcommand{\Cxkp}{\mC_{k}^{p}}             
\newcommand{\Cxke}{\mC_{k}^{e}}             
\newcommand{\Cxkke}{\mC_{k-1}^{e}}          

\newcommand{\vy}{\vec{y}}                   
\newcommand{\vyk}{\vy_{k}}                  

\newcommand{\evy}{\mean{\vy}}               
\newcommand{\evyk}{\evy_{k}}                

\newcommand{\Cy}{\mC^{y}}                   
\newcommand{\Cyk}{\Cy_{k}}                  

\newcommand{\Cxy}{\mC^{x,y}}                
\newcommand{\Cxyk}{\Cxy_{k}}                

\newcommand{\meas}{\tilde{y}}               

\newcommand{\vmeas}{\tilde{\vec{y}}}        
\newcommand{\vmeask}{\vmeas_{k}}            
\newcommand{\vmeaskk}{\vmeas_{k-1}}         

\newcommand{\vmeasks}{\vmeas_{k:1}}         
\newcommand{\vmeaskks}{\vmeas_{k-1:1}}      

\newcommand{\measset}{\mathcal{Y}}
\newcommand{\meassetk}{\measset_{k}}

\newcommand{\vv}{\vec{v}}                   
\newcommand{\vvk}{\vv_{k}}                  

\newcommand{\evv}{\mean{\vv}}               
\newcommand{\evvk}{\evv_{k}}                

\newcommand{\Cv}{\mC^{v}}                   
\newcommand{\Cvk}{\Cv_{k}}                  

\newcommand{\vw}{\vec{w}}                   
\newcommand{\vwk}{\vw_{k}}                  

\newcommand{\evw}{\mean{\vw}}               
\newcommand{\evwk}{\evw_{k}}                

\newcommand{\Cw}{\mC^{w}}                   
\newcommand{\Cwk}{\Cw_{k}}                  

\newcommand{\fp}{f^{p}}                     
\newcommand{\fpk}{\fp_{k}}                  

\newcommand{\fe}{f^{e}}                     
\newcommand{\fek}{\fe_{k}}                  
\newcommand{\fekk}{\fe_{k-1}}               

\newcommand{\fa}{f^{a}}                     
\newcommand{\fak}{\fa_{k}}                  

\newcommand{\fh}{f^{h}}                     
\newcommand{\fhk}{\fh_{k}}                  

\newcommand{\fy}{f^{y}}                     
\newcommand{\fyk}{\fy_{k}}                  

\newcommand{\fv}{f^{v}}                     
\newcommand{\fvk}{\fv_{k}}                  

\newcommand{\fw}{f^{w}}                     
\newcommand{\fwk}{\fw_{k}}                  

\newcommand{\fxyk}{f_{k}^{x,y}}             

\newcommand{\alphaki}{\alpha_{k,i}}

\newcommand{\vxi}{\vx_{i}}

\newcommand{\vxki}{\vx_{k,i}}
\newcommand{\vxkki}{\vx_{k-1,i}}

\newcommand{\vs}{\vec{s}}

\newcommand{\vsi}{\vs_{i}}
\newcommand{\vsj}{\vs_{j}}
\newcommand{\vsk}{\vs_{k}}
\newcommand{\vski}{\vs_{k,i}}

\newcommand{\vwki}{\vw_{k,i}}
\newcommand{\vvki}{\vv_{k,i}}

\newcommand{\vzi}{\vz_{i}}

\newcommand{\Cz}{\mC^{z}}

\newcommand{\va}{\vec{a}}                   
\newcommand{\vak}{\va_{k}}                  

\newcommand{\vh}{\vec{h}}                   
\newcommand{\vhk}{\vh_{k}}                  

\newcommand{\Bmat}{\begin{bmatrix}}
\newcommand{\Emat}{\end{bmatrix}}

\newcommand{\Beq}{\begin{equation}\begin{aligned}}
\newcommand{\Eeq}{\end{aligned}\end{equation}}

\newcommand{\Sec}[1]{Sec.~\ref{#1}}             
\newcommand{\App}[1]{\ref{#1}}                  
\newcommand{\Eq}[1]{\eqref{#1}}                 
\newcommand{\Fig}[1]{Fig.~\ref{#1}}             
\newcommand{\Tab}[1]{Table~\ref{#1}}            

\makeatletter
\newcounter{ALC@tempcntr}                       

\makeatother

\theoremstyle{plain}

\newtheorem{Theorem}   {Theorem}   [section]
\newtheorem{Definition}{Definition}[section]

\begin{document}
    \begin{frontmatter}
        \title{The Smart Sampling Kalman Filter\\with Symmetric Samples}
        
        \author{Jannik~Steinbring}
        \ead{jannik.steinbring@kit.edu}
        
        \author{Martin~Pander}
        \ead{martin.pander@student.kit.edu}
        
        \author{Uwe~D.~Hanebeck}
        \ead{uwe.hanebeck@ieee.org}
        
        \address{Intelligent Sensor-Actuator-Systems Laboratory (ISAS)\\
                 Institute for Anthropomatics and Robotics\\
                 Karlsruhe Institute of Technology (KIT), Germany\vspace{3mm}}

\begin{abstract}

Nonlinear Kalman Filters are powerful and widely-used techniques when trying to estimate the hidden state of a stochastic nonlinear dynamic system.
In this paper, we extend the Smart Sampling Kalman Filter~(\sskf) with a new point symmetric Gaussian sampling scheme.
This not only improves the \sskf's estimation quality, but also reduces the time needed to compute the required optimal Gaussian samples drastically.
Moreover, we improve the numerical stability of the sample computation, which allows us to accurately approximate a thousand-dimensional Gaussian distribution using tens of thousands of optimally placed samples.
We evaluate the new symmetric \sskf by computing higher-order moments of standard normal distributions and investigate the estimation quality of the \sskf when dealing with symmetric measurement equations.
Finally, extended object tracking based on many measurements per time step is considered.
This high-dimensional estimation problem shows the advantage of the \sskf being able to use an arbitrary number of samples independent of the state dimension, in contrast to other state-of-the-art sample-based Kalman Filters.

\end{abstract}

    \end{frontmatter}

\section{Introduction}
\label{sec:introduction}

Estimating the hidden state of a stochastic dynamic system based on noisy measurements is crucial for many applications in control, object tracking, or robotics.
When considering linear systems corrupted by additive Gaussian noise, the Kalman Filter (KF) is the optimal estimator with respect to the mean square error \cite{rudolf_e._kalman_new_1960}.
Unfortunately, most practical problems are nonlinear, making closed-form solutions intractable.
Consequently, approximative approaches have to be used.
Particle Filters (PFs) \cite{sanjeev_arulampalam_tutorial_2002, branko_ristic_beyond_2004, arnaud_doucet_tutorial_2011, jayesh_h._kotecha_gaussian_2003} try to approximate the complete, in general multimodal, system state density with a set of weighted particles.
This comes at the cost of computational complexity due to the curse of dimensionality.
Another problem is sample degeneracy, in particular for high-dimensional state spaces, as a consequence of the particle reweighting using the likelihood function.
To reduce computational complexity and circumvent the problem of sample degeneracy, the Progressive Gaussian Filter (PGF)~\cite{jannik_steinbring_progressive_2014} approximates the system state as a Gaussian and moves the particles automatically to the important regions of the state space.
Nevertheless, those nonlinear filters are still costly compared to linear filters applied to nonlinear problems.

The Extended Kalman Filter (EKF) explicitly linearizes the underlying models around the current state estimate to be able to apply the standard KF to the considered problem~\cite{dan_simon_optimal_2006}.
Iterated variants of the EKF (IEKF) try to improve the EKF approach by iteratively searching for a more suitable point for the model linearization~\cite{dan_simon_optimal_2006}.
A more suitable way of model linearization is based on statistical linearization, which can be performed in the best case analytically or, in all other cases, by exploiting samples in the form of Linear Regression Kalman Filters (LRKFs)~\cite{tine_lefebvre_kalman_2004}.
LRKFs obtain the required moments by propagating samples through the system and measurement models and computing sample mean and sample covariance matrix, respectively.
The most commonly used LRKF is the Unscented Kalman Filter (UKF)~\cite{simon_j._julier_unscented_2004}.
Its samples are, however, limited in number and placement, and several attempts exist to improve the UKF by finding its optimal parameter settings for specific estimation problems~\cite{ondrej_straka_unscented_2014}.
Nevertheless, the additional computational time required to find proper UKF parameters can be used instead to propagate more carefully chosen samples through the models in order to improve the estimation quality.
For example, the Gauss-Hermite Kalman Filter (GHKF) introduced in \cite{kazufumi_ito_gaussian_2000} is based on the Gauss-Hermite quadrature rule to generate its samples.
Unfortunately, the GHKF also suffers from the curse of dimensionality, and hence, is not well suited for larger state spaces.
The fifth-degree Cubature Kalman Filter~(CKF) \cite{bin_jia_high-degree_2013} relies on a fifth-degree spherical-radial integration rule to construct its samples.
However, by design, the number of samples still grows quadratically in the state dimension making the \ckf computational burdensome when dealing with larger state spaces.
A non-deterministic LRKF was proposed with the Randomized UKF (RUKF)~\cite{jindrich_dunik_development_2011, jindrich_dunik_stochastic_2013}.
Here, an arbitrary number of randomly scaled and rotated UKF sample sets are combined to a single set of samples.
On the one hand this has the advantage of being able to change the employed number of samples.
On the other hand it prohibits a reproducible filter behavior and imposes an additional runtime overhead compared to other LRKFs due to the creation of several random orthogonal matrices per prediction and measurement update.
The estimation quality of any LRKF, regardless of the sampling it is based on, can be improved by using the iterated statistical linearization approach~\cite{angel_f._garcia-fernandez_iterated_2014}.
A more detailed overview of linear filters and LRKFs can be found in~\cite{pawe_stano_parametric_2013, jannik_steinbring_lrkf_2014}.

Recently, the Smart Sampling Kalman Filter~(\sskf) was proposed in \cite{jannik_steinbring_s2kf:_2013, jannik_steinbring_lrkf_2014}, and already successfully used for Simultaneous Localization and Mapping (SLAM) in~\cite{cihan_ulas_planar_2014}.
The \sskf uses optimal deterministic sampling of a standard normal distribution comprising an arbitrary number of samples based on a combination of the Localized Cumulative Distribution (LCD) and a modified Cramér-von Mises distance \cite{uwe_d._hanebeck_dirac_2009, igor_gilitschenski_efficient_2013}.
The same LCD approach was also extended to approximate arbitrary Gaussian mixture distributions \cite{igor_gilitschenski_deterministic_2014}.
In this paper, we extend the \sskf with point symmetric Gaussian sampling.
This new symmetric sampling reflects the point symmetry of the Gaussian distribution and allows for matching all odd moments of a standard normal distribution exactly, which results in a more accurate state estimation.
We also improve the numerical stability of the LCD approach.
As a result, it is now possible to compute an optimal approximation of a thousand-dimensional standard normal distribution comprising tens of thousands of samples.
Moreover, due to the exploited point symmetry, the required number of samples that have to be optimized is reduced by half.
Consequently, the new samples can be computed much faster.
In this regard, the \sskf catches up to state-of-the-art LRKFs as all of them also rely on a point symmetric sampling.

The remainder of the paper is organized as follows.
First, we give a detailed problem formulation including the general recursive Bayesian estimation, its transition to Nonlinear Kalman Filters, and finally their approximation using LRKFs.
After that, in \Sec{sec:symmetric-sampling}, we introduce a new point symmetric version of the \sskf.
In \Sec{sec:evaluation}, we evaluate the new symmetric \sskf by computing higher-order moments of multivariate standard normal distributions, showing the advantage of the new point symmetric sampling scheme when dealing with symmetric measurement equations, and performing extended object tracking.
Finally, conclusions are given in \Sec{sec:conclusions}.

\section{Problem Formulation}
\label{sec:problem-formulation}

We consider estimating the hidden state $\vxk$ of a discrete-time stochastic nonlinear dynamic system, where the system model
\Beq
    \label{eq:sys-model}
    \vxk = \vak(\vxkk, \vwk)
\Eeq
describes its temporal evolution\footnote{The subscript $k$ denotes the discrete time step and vectors are underlined.}.
Additionally, we receive noisy measurements $\vmeask$ that are assumed to be generated according to the measurement model
\Beq
    \label{eq:meas-model}
    \vyk = \vhk(\vxk, \vvk) \enspace.
\Eeq
Thus, the received measurements $\vmeask$ are realizations of the random variable $\vyk$.
The noise variables $\vwk$ and $\vvk$ are assumed to be Gaussian and independent of the system state for all time steps.

Our goal is to determine a state estimate of $\vxk$ in the form of a conditional state density
\Beq
    \label{eq:filtered-state}
    \fek(\vxk) :=  f(\vxk \cond \vmeasks) = f(\vxk \cond \vmeask, \ldots, \vmeas_1)
\Eeq
recursively over time using Bayesian inference.
Such a recursive estimator consists of two parts, namely the prediction step and the filter step.
On the one hand, the prediction step propagates the state estimate $\fekk(\vxkk)$ from time step~$k - 1$ to the current time step~$k$ by employing the system model \Eq{eq:sys-model} resulting in the predicted state density
\Beq
    \label{eq:predicted-state}
    \fpk(\vxk) := f(\vxk \cond \vmeaskks) = f(\vxk \cond \vmeaskk, \ldots, \vmeas_1) \enspace.
\Eeq
On the other hand, the filter step incorporates a newly received measurement $\vmeask$ into this propagated state estimate $\fpk(\vxk)$ with the aid of the measurement model \Eq{eq:meas-model}.

To compute the prediction step, \Eq{eq:sys-model} has to be transformed into a state transition density according to
\Beq
    \fak(\vxk \cond \vxkk) := \int \delta(\vxk - \vak(\vxkk, \vwk)) \cdot \fwk(\vwk) \dd \vwk \enspace,
\Eeq
where $\delta(\cdot)$ denotes the Dirac delta distribution and $\fwk(\cdot)$ the system noise density
\Beq
    \fwk(\vwk) = \Gauss(\vwk; \evwk, \Cwk) \enspace.
\Eeq
Based on this, the predicted state density can be obtained with the Chapman-Kolomogorov equation~\cite{yaakov_bar-shalom_estimation_2001}
\Beq
    \label{eq:time-update}
    \fpk(\vxk) &= \int \fak(\vxk \cond \vxkk) \cdot \fekk(\vxkk) \dd \vxkk \\
               &= \iint \delta(\vxk - \vak(\vxkk, \vwk)) \cdot \fekk(\vxkk) \cdot \fwk(\vwk) \dd \vxkk \dd \vwk \enspace.
\Eeq

Concerning the filter step, Bayes' rule serves as its fundamental basis and the posterior, i.e., filtered, state density is obtained according to
\Beq
    \label{eq:measurement-update}
    \fek(\vxk) = \frac{\fhk(\vmeask \cond \vxk) \cdot \fpk(\vxk)}{\fyk(\vmeask \cond \vmeaskks)} \enspace,
\Eeq
where $\fhk(\cdot)$ denotes the likelihood function and $\fyk(\cdot)$ the measurement distribution given all measurements up to the time step $k - 1$.
As $\fyk(\cdot)$ is independent of $\vxk$, it is only a normalization constant assuring that $\fek(\vxk)$ is a valid density function.
The required likelihood function is obtained in a similar manner to the state transition density according to
\Beq
    \label{eq:likelihood}
    \fhk(\vmeask \cond \vxk) := \int \delta(\vmeask - \vhk(\vxk, \vvk)) \cdot \fvk(\vvk) \dd \vvk \enspace,
\Eeq
with the measurement noise density
\Beq
    \fvk(\vvk) = \Gauss(\vvk; \evvk \Cvk) \enspace.
\Eeq

Given an initial state estimate $\fe_0(\vx_0)$, the recursive state estimation of $\fek(\vxk)$ is performed with the alternate use of the prediction step~\Eq{eq:time-update} and the filter step~\Eq{eq:measurement-update}.
However, computing both steps analytically is almost always impossible except for special cases such as linear models corrupted by additive Gaussian noise.
Consequently, approximative solutions of the recursive Bayesian estimation have to be used instead.

\subsection{Nonlinear Kalman Filters}
\label{sec:nonlinear-kfs}

Besides the popular Particle Filters, an important approximative Bayesian estimation technique is given by the class of Nonlinear Kalman Filters.
These filters make two simplifications of the general Bayesian estimator described above.

First, the filtered as well as the predicted state estimates are always approximated as Gaussians.
Hence, the predicted state density is given by
\Beq
    \label{eq:predicted-state-gaussian}
    \fpk(\vxk) \approx \Gauss(\vxk; \evxkp, \Cxkp) \enspace,
\Eeq
with predicted state mean
\Beq
    \label{eq:time-update-mean}
    \evxkp &= \int \vxk \cdot \fpk(\vxk) \dd \vxk \\
           &= \iint \vak(\vxkk, \vwk) \cdot \fekk(\vxkk) \cdot \fwk(\vwk) \dd \vxkk \dd \vwk
\Eeq
and predicted state covariance matrix
\Beq
    \label{eq:time-update-cov}
    \Cxkp &= \int (\vxk - \evxkp) \cdot (\vxk - \evxkp)\T \cdot \fpk(\vxk) \dd \vxk \\
          &= \iint (\vak(\vxkk, \vwk) - \evxkp) \cdot (\vak(\vxkk, \vwk) - \evxkp)\T \cdot \fekk(\vxkk) \cdot \fwk(\vwk) \dd \vxkk \dd \vwk \enspace,
\Eeq
respectively.

Second, the Bayesian filter step \Eq{eq:measurement-update} can be reformulated in form of the joint density $f_k^{x,y}(\vxk, \vyk)$ of predicted state $\vxk$ and measurement $\vyk$ according to
\Beq
    \fek(\vxk) = \frac{f_k^{x,y}(\vxk, \vmeask \cond \vmeaskks)}{\fyk(\vmeask \cond \vmeaskks)} \enspace.
\Eeq
In Kalman filtering, this joint density is also approximated as a Gaussian resulting in the approximative Gaussian posterior state density
\Beq
    \fek(\vxk) &\approx \frac{\JointGauss{\vxk}{\vmeask}{\evxkp}{\evyk}{\Cxkp}{\Cxyk}{(\Cxyk)\T}{\Cyk}}
                             {\fyk(\vmeask \cond \vmeaskks)} = \Gauss(\vxk; \evxke, \Cxke) \enspace,
\Eeq
with posterior state mean
\Beq
    \label{eq:linear-meas-update-mean}
    \evxke = \evxkp + \Cxyk \cdot \inv{\Cyk} \cdot (\vmeask - \evyk)
\Eeq
and posterior state covariance matrix
\Beq
    \label{eq:linear-meas-update-cov}
    \Cxke = \Cxkp - \Cxyk \cdot \inv{\Cyk} \cdot (\Cxyk)\T \enspace,
\Eeq
which are the well-known Kalman Filter formulas\cite{dan_simon_optimal_2006}.
In order to obtain \Eq{eq:linear-meas-update-mean} and \Eq{eq:linear-meas-update-cov}, the measurement mean
\Beq
    \label{eq:meas-mean}
    \evyk &= \int \vyk \cdot \fyk(\vyk) \; \dd \vyk \\
          &= \iint \vhk(\vxk, \vvk) \cdot \fpk(\vxk) \cdot \fvk(\vvk) \; \dd \vxk \dd \vvk \enspace,
\Eeq
the measurement covariance matrix
\Beq
    \label{eq:meas-cov}
    \Cyk &= \int (\vyk - \evyk) \cdot (\vyk - \evyk)\T \cdot \fyk(\vyk) \; \dd \vyk \\
         &= \iint (\vhk(\vxk, \vvk) - \evyk) \cdot (\vhk(\vxk, \vvk) - \evyk)\T \cdot \fpk(\vxk) \cdot \fvk(\vvk) \; \dd \vxk \dd \vvk \enspace,
\Eeq
as well as the cross-covariance matrix of predicted state and measurement
\Beq
	\label{eq:state-meas-cross-cov}
	\Cxyk &= \iint (\vxk - \evxkp) \cdot (\vyk - \evyk)\T \cdot \fxyk(\vxk, \vyk) \; \dd \vxk \dd \vyk \\
	      &= \iint (\vxk - \evxkp) \cdot (\vhk(\vxk, \vvk) - \evyk)\T \cdot \fpk(\vxk) \cdot \fvk(\vvk) \; \dd \vxk \dd \vvk
\Eeq
are required.

In summary, running a Nonlinear Kalman Filter boils down to computing two integrals for the prediction step and three integrals for the filter step.
Moreover, note that no explicit likelihood function \Eq{eq:likelihood} is required at all, which eases filter development.

\subsection{Linear Regression Kalman Filters}
\label{sec:lrkfs}

Unfortunately, computing the above integrals in closed-form is only possible for a small set of system and measurement models.
This includes for example polynomials, trigonometric and, of course, linear functions (leading to the classical linear Kalman Filter).
The closed-from integration yields the best possible Kalman Filter for the given models.
In all other cases, numerical integration methods have to be applied, which can result in a reduced estimation performance and an increased computational demand.

As we aim for an online estimation technique, the employed integration method has to possess a \emph{real-time capable computational complexity} and still deliver adequate integration results in order to obtain a good recursive state estimation quality.
When looking at the five integrals in \Sec{sec:nonlinear-kfs}, it can be seen that the last terms are always a product of two independent Gaussian densities, namely
\Beq
    \label{eq:state-sys-noise-pdf}
    \fekk(\vxkk) \cdot \fwk(\vwk) = \JointGauss{\vxkk}{\vwk}{\evxkke}{\evwk}{\Cxkke}{\mat{0}}{\mat{0}}{\Cwk}
\Eeq
for the prediction and
\Beq
    \label{eq:state-meas-noise-pdf}
    \fpk(\vxk) \cdot \fvk(\vvk) = \JointGauss{\vxk}{\vvk}{\evxkp}{\evvk}{\Cxkp}{\mat{0}}{\mat{0}}{\Cvk}
\Eeq
for the measurement update, respectively.

By exploiting this fact, an efficient, i.e., fast but still accurate, computation of the integrals is possible.
This can be done by replacing the occurring Gaussian distributions \Eq{eq:state-sys-noise-pdf} and \Eq{eq:state-meas-noise-pdf} with proper Dirac mixture densities, that is, sample-based density representations, and evaluating the system model \Eq{eq:sys-model} and measurement model \Eq{eq:meas-model} using these samples.
As a result, emphasis is directly put on the important regions of the state space, and the regions covered by only a small portion of the probability mass of the Gaussian densities are neglected.
This approach leads to the class of Linear Regression Kalman Filters (LRKFs).

A Dirac mixture approximation of a given probability density function $f_k(\vsk)$ comprising $M_k$ samples with sample positions $\vski$ and sample weights $\alphaki$ is defined as
\Beq
    \label{eq:dirac-mixture}
    \sum_{i=1}^{M_k} \alphaki \cdot \delta(\vsk - \vski) \enspace,
\Eeq
where for the sample weights
\Beq
    \sum_{i = 1}^{M_k} \alphaki = 1
\Eeq
holds.
Such an approximation can be computed in several ways, e.g., by simply using random sampling or deterministic approaches such as done by the UKF.

Now, we assume that an approximation
\Beq
    \label{eq:state-sys-noise-approx}
    \sum_{i=1}^{M_k} \alphaki \cdot \delta\left(\Bmat \vxkk \\ \vwk \Emat - \Bmat \vxkki \\ \vwki \Emat\right) \enspace
\Eeq
of the Gaussian joint density \Eq{eq:state-sys-noise-pdf} comprising $M_k$ samples with positions $[\vxkki\T,\;\vwki\T]\T$ and weights $\alphaki$ is at hand.
By replacing the Gaussian joint density in the integrals \Eq{eq:time-update-mean} and \Eq{eq:time-update-cov} with the Dirac mixture approximation \Eq{eq:state-sys-noise-approx}, and using the Dirac sifting property, we obtain an approximation for the predicted state mean
\Beq
    \label{eq:time-update-sample-mean}
    \evxkp \approx \sum_{i=1}^{M_k} \alphaki \cdot \vak(\vxkki, \vwki)
\Eeq
and the predicted state covariance matrix
\Beq
    \label{eq:time-update-sample-cov}
    \Cxkp &\approx \sum_{i=1}^{M_k} \alphaki \cdot (\vak(\vxkki, \vwki) - \evxkp) \cdot (\vak(\vxkki, \vwki) - \evxkp)\T \enspace.
\Eeq
The same procedure is used for computing the integrals required for the measurement update.
First, a Dirac mixture approximation
\Beq
    \label{eq:state-meas-noise-approx}
        \sum_{i=1}^{M_k} \alphaki \cdot \delta\left(\Bmat \vxk \\ \vvk \Emat - \Bmat \vxki \\ \vvki \Emat\right) \enspace,
\Eeq
of the Gaussian \Eq{eq:state-meas-noise-pdf} encompassing $M_k$ samples with positions $[\vxki\T,\;\vvki\T]\T$ and weights $\alphaki$ is computed.
Second, by replacing the joint Gaussian with its Dirac mixture approximation in the three integrals \Eq{eq:meas-mean}, \Eq{eq:meas-cov}, and \Eq{eq:state-meas-cross-cov}, and using once more the Dirac sifting property, we get an approximation for the measurement mean
\Beq
    \label{eq:meas-sample-mean}
    \evyk \approx \sum_{i=1}^{M_k} \alphaki \cdot \vhk(\vxki, \vvki) \enspace,
\Eeq
the measurement covariance matrix
\Beq
    \label{eq:meas-sample-cov}
    \Cyk &\approx \sum_{i=1}^{M_k} \alphaki \cdot (\vhk(\vxki, \vvki) - \evyk) \cdot (\vhk(\vxki, \vvki) - \evyk)\T \enspace,
\Eeq
and the cross-covariance matrix
\Beq
    \label{eq:state-meas-sample-cross-cov}
	\Cxyk \approx \sum_{i=1}^{M_k} \alphaki \cdot (\vxki -  \evxkp) \cdot (\vhk(\vxki, \vvki) - \evyk)\T \enspace.
\Eeq
It should be noted that the number of samples for the time and the measurement update do not have to be the same.
Moreover, the Dirac mixture approximations \Eq{eq:state-sys-noise-approx} and \Eq{eq:state-meas-noise-approx} can be completely different in the way they are obtained, although this is usually not the case.

\section{The Smart Sampling Kalman Filter with Symmetric Samples}
\label{sec:symmetric-sampling}

In \cite{uwe_d._hanebeck_dirac_2009}, the authors proposed an approach based on the Localized Cumulative Distribution (LCD) to optimally approximate Gaussian distributions with a set of equally weighted samples.
This is done by transforming the approximation problem into an optimization problem.
Unfortunately, such optimization is very time-consuming, and hence, not suitable for online nonlinear filtering.
To enable the LCD approach for online filtering, it is used to optimally sample only a standard normal distribution offline (before filter usage) and transform these samples online (during filter usage) to any required Gaussian with the aid of the Mahalanobis transformation \cite{wolfgang_hardle_applied_2008}.
This is the fundamental basis for the \sskf \cite{jannik_steinbring_lrkf_2014}.
But other nonlinear estimators such as the Progressive Gaussian Filter also make use of this sampling technique.

However, the current LCD approach can, and will, arrange the samples in an arbitrary way to optimally approximate a standard normal distribution.
More precisely, it does not take the point symmetry of the standard normal distribution explicitly into account.
This can lead to a set of asymmetric samples.
Here, we extend the LCD approach to approximate an $N$-dimensional standard normal distribution with a set of \emph{point symmetric} and equally weighted samples.
Moreover, we improve the numerical stability of the LCD approach to allow approximations of very high dimensions.
This new optimal point symmetric sampling is then used to obtain a symmetric version of the \sskf.

The use of point symmetric samples offers several benefits.
First, the symmetric samples reflect the symmetry of the standard normal distribution allowing for more accurate estimation results as will be seen in the evaluation.
Second, the used point symmetry makes it possible to capture \emph{all odd moments} of the standard normal distribution exactly (a proof is given in \App{app:proof-odd-moments}).
Finally, the required number of sample positions that have to be optimized is reduced by half.
Consequently, an approximation can be computed much faster.
Besides point symmetry, other symmetries such as axial symmetry could also be exploited.
However, this would prevent us from using an arbitrary number of samples and would limit the optimizer's control over the sample placement.

In the following, we first define the set of parameters describing point symmetric Dirac mixtures in \Sec{sec:sym-dirac-mixtures}.
These parameters have then to be optimized in order to approximate a standard normal distribution in an optimal way.
This requires distance measures between a standard normal distribution and the point symmetric Dirac mixtures given in \Sec{sec:distance-measures}.
Subsequently, the gradients of the distance measures are derived in \Sec{sec:distance-measure-gradients}.
Finally, in \Sec{sec:sym-s2kf}, we give a procedure to compute point symmetric Dirac mixture approximations of standard normal distributions based on the introduced distance measures and their gradients.

\subsection{Point Symmetric Dirac Mixtures}
\label{sec:sym-dirac-mixtures}

First, we have to modify the generic Dirac mixture \Eq{eq:dirac-mixture} to obtain a symmetric one.
This is performed by distinguishing between an even and odd number of samples.
For the case of $2L$ samples with $L \in \IN_+$, that is, the even case, we place the samples point symmetrically around the state space origin yielding the equally weighted Dirac mixture
\Beq
    \label{eq:sym-dirac-mixture-even}
    \frac{1}{2L} \sum_{i=1}^L \delta(\vs - \vsi) + \delta(\vs + \vsi) \enspace,
\Eeq
with sample positions $\vsi$ and $-\vsi$.
For $2L + 1$ samples, the odd case, we additionally place a sample fixed at the state space origin and obtain the Dirac mixture
\Beq
    \label{eq:sym-dirac-mixture-odd}
    \frac{1}{2L + 1} \left(\delta(\vs) + \sum_{i=1}^L \delta(\vs - \vsi) + \delta(\vs + \vsi) \right) \enspace.
\Eeq
This preserves the desired point symmetry.
As the position of the additional sample in the odd case is constant, the set of parameters
\Beq
    S := \{\vs_1, \dots, \vs_L\}
\Eeq
is the same for both Dirac mixtures.

Note that the UKF sample set (with equally weighted samples) is a special case of these point symmetric Dirac mixtures \cite{simon_j._julier_unscented_2004}.
With an even number of samples, it has the parametrization
\Beq
    \vsi = \sqrt{N} \cdot \ve_i \quad \forall i \in \{1, \ldots, N\} \enspace,
\Eeq
where $\ve_i$ denotes the unit vector along the $i$-th dimension.
In the odd case, the parametrization is
\Beq
    \vsi = \sqrt{N + 0.5} \cdot \ve_i \quad \forall i \in \{1, \ldots, N\} \enspace,
\Eeq
that is, the sample spread is larger due to the additional point mass at the state space origin.

\subsection{Distance Measures}
\label{sec:distance-measures}

Our goal is to determine the set of parameters $S$ for the above Dirac mixtures so that they approximate a multivariate standard normal distribution in an optimal way.
This requires a distance measure between the involved continuous and discrete distributions.
As the classical cumulative distribution function is not suitable for the multi-dimensional case \cite{uwe_d._hanebeck_localized_2008}, we utilize the LCD approach in the same way as the asymmetric \sskf.
\begin{Definition}[Localized Cumulative Distribution \cite{jannik_steinbring_lrkf_2014}]\hfill\\
    Let $f(\vs)$ be a $N$-dimensional density function.
    The corresponding Localized Cumulative Distribution is defined as
    \Beq
        \label{eq:lcd}
        F(\vm, b) = \int_{\IR^N} f(\vs) \cdot K(\vs - \vm, b) \dd \vs \enspace,
    \Eeq
    with $\vm \in \IR^N$, $b \in \IRplus$, and the symmetric and integrable kernel
    \Beq
        \label{eq:lcd-kernel}
        K(\vs - \vm, b) = \gaussExp{\sqNorm{\vs - \vm}}{b^2} \enspace.
    \Eeq
    Here, $\vm$ characterizes the location of the kernel and $b$ its size.
\end{Definition}

The LCD of an $N$-dimensional standard normal distribution is an integral of a product of two (unnormalized) Gaussians.
By using the fact that the product of two Gaussian distributions is also an unnormalized Gaussian and the integral over a probability density equals one, its LCD is obtained by~\cite{uwe_d._hanebeck_dirac_2009}
\Beq
    \label{eq:lcd-std-normal}
    \LCDN(\vm, b) &= \int_{\IR^N} \Gauss(\vs\,; \vzero, \mI_N) \cdot (2\pi)^{\frac{N}{2}} b^N
                                  \Gauss(\vs\,; \vm, b^2\mI_N) \dd \vs \\
                  &= \frac{(2\pi)^{\frac{N}{2}} b^N}{(2\pi)^{\frac{N}{2}}\sqrt{|(1+b^2)\mI_N|}} 
                     \gaussExp{\sqNorm{\vm}}{(1 + b^2)} \\
                  &= \left(\frac{b^2}{1 + b^2}\right)^{\hspace*{-0.1cm}\frac{N}{2}}
                     \gaussExp{\sqNorm{\vm}}{(1 + b^2)} \enspace,
\Eeq
where $\mI_N$ denotes the identity matrix of dimension $N$.
Based on the Dirac sifting property, the LCD of the Dirac mixture comprising an even number of samples is given by
\Beq
    \label{eq:lcd-dm-even}
    \LCDDMEven(S, \vm, b) &= \frac{1}{2L} \left(\sum_{i=1}^L \gaussExp{\sqNorm{\vsi - \vm}}{b^2} + \gaussExp{\sqNorm{-\vsi - \vm}}{b^2}\right) \enspace,
\Eeq
whereas the LCD of the odd Dirac mixture is
\Beq
    \label{eq:lcd-dm-odd}
    \LCDDMOdd(S, \vm, b) &= \frac{1}{2L + 1} \left(\gaussExp{\sqNorm{\vm}}{b^2} + \sum_{i=1}^L \gaussExp{\sqNorm{\vsi - \vm}}{b^2} + \gaussExp{\sqNorm{-\vsi - \vm}}{b^2}\right) \,. 
\Eeq
To compare the standard normal LCD with a Dirac mixture LCD, we use the modified Cramér-von Mises distance defined as follows.

\begin{Definition}[Modified Cramér–von Mises Distance]\hfill\\
    The modified Cramér–von Mises (CvM) distance $D$ between two LCDs $F(\vm, b)$ and $\tilde{F}(\vm, b)$ is given by
    \Beq
        \label{eq:mcvm-dist}
        D(F, \tilde{F}) = \int_0^\infty w(b) \int_{\IR^N} \left(F(\vm, b) - \tilde{F}(\vm, b)\right)^2 \dd \vm \; \dd b \enspace,
    \Eeq
    with weighting function
    \Beq
        \label{eq:weighting-func}
        w(b) = \begin{cases}
                \pi^{-\frac{N}{2}} b^{1 - N} & \text{, } b \in [0, \bMax] \\
                0 & \text{, elsewhere} \enspace.
               \end{cases}
    \Eeq
\end{Definition}

The new term $\pi^{-\frac{N}{2}}$ in the weighting function $w(b)$ (in contrast with the definition in \cite{jannik_steinbring_lrkf_2014}) is a consequence of the involved LCDs $\LCDN$, $\LCDDMEven$, and $\LCDDMOdd$.
Without this term, the modified CvM distances between these LCDs would be unbounded for an increasing dimension $N$, which in turn would make the distances numerically unstable.
This improvement now allows the \sskf to compute Dirac mixture approximations for very high state dimensions, e.g., $N > 200$.

First, we consider the distance between the standard normal distribution and the Dirac mixture comprising an even number of samples, and then extend the results to the odd case.
The distance~$D(\LCDN, \LCDDMEven)$ can be split into three terms according to
\Beq
    \label{eq:mcvm-dist-even}
    D(\LCDN, \LCDDMEven) = D^e(S) = D^e_1 - 2 D^e_2(S) + D^e_3(S) \enspace,
\Eeq
with the sample-independent part
\Beq
    \label{eq:mcvm-dist-even-d1}
    D^e_1 &= \int_0^{\bMax} b \left(\frac{b^2}{1 + b^2}\right)^{\hspace*{-0.1cm}\frac{N}{2}} \dd b \enspace,
\Eeq
and the sample-dependent terms
\Beq
    \label{eq:mcvm-dist-even-d2}
    D^e_2(S) &= \int_0^{\bMax} \frac{2b}{2L} \left(\frac{2 b^2}{1 + 2 b^2}\right)^{\hspace*{-0.1cm}\frac{N}{2}} \cdot \sum_{i=1}^L \gaussExp{\sqNorm{\vsi}}{(1 + 2b^2)} \dd b \enspace,
\Eeq
and
\Beq
    \label{eq:mcvm-dist-even-d3}
    D^e_3(S) &= \int_0^{\bMax} \hspace*{-0.1cm} \frac{2b}{(2L)^2} \sum_{i=1}^L \sum_{j=1}^L \gaussExp{\sqNorm{\vsi - \vsj}}{2b^2} + \gaussExp{\sqNorm{\vsi + \vsj}}{2b^2} \dd b \enspace.
\Eeq
The proof is given in \App{app:proof-dist-even}.
Note that the integration over $b$ is bounded by $\bMax$ due to the support of the weighting function~$w(b)$.
To speed up the distance computation, the following theorem can be applied.
\begin{Theorem}
\label{theorem-closed-form-even-d3}\hfill\\
    For a given $\bMax$, the following expression for $D^e_3(S)$ can be obtained
    \Beq
        \label{eq:closed-form-even-d3}
        D^e_3(S) &= \frac{2}{(2L)^2} \sum_{i=1}^L \sum_{j=1}^L
            \frac{\bMax^2}{2}
            \left( \gaussExp{\sqNorm{\vsi - \vsj}}{2\bMax^2} + \gaussExp{\sqNorm{\vsi + \vsj}}{2\bMax^2} \right)\,+ \\
            &\hspace*{2.8cm} \frac{1}{8} \left(\sqNorm{\vsi - \vsj} \Ei\left(-\frac{1}{2}\frac{\sqNorm{\vsi - \vsj}}{2\bMax^2}\right) + \sqNorm{\vsi + \vsj} \Ei\left(-\frac{1}{2}\frac{\sqNorm{\vsi + \vsj}}{2\bMax^2}\right)\right) \enspace,
    \Eeq
    where $\Ei(\cdot)$ denotes the exponential integral.
    \begin{proof}
        The proof is given in \App{app:proof-closed-form-even-d3}.
    \end{proof}
\end{Theorem}

Now, we consider the case of an odd number of samples.
Like in the even case, $D(\LCDN, \LCDDMOdd)$ can be split into three terms
\Beq
    \label{eq:mcvm-dist-odd}
    D(\LCDN, \LCDDMOdd) = D^o(S) = D^o_1 - 2 D^o_2(S) + D^o_3(S) \enspace.
\Eeq
The first part $D^o_1$ is also independent of the samples $S$ and identical to its even counterpart, i.e.,
\Beq
    \label{eq:mcvm-dist-odd-d1}
    D^o_1 &= D^e_1 \enspace.
\Eeq
The sample-dependent terms $D^o_2(S)$ and $D^o_3(S)$ can be expressed in terms of the even case plus additional terms due to the fixed sample at the state space origin according to
\Beq
    \label{eq:mcvm-dist-odd-d2}
    D^o_2(S) &= \frac{2L}{2L + 1} D^e_2(S) +
                \int_0^{\bMax} \hspace*{-0.1cm} \frac{b}{2L + 1}
                \left(\frac{2 b^2}{1 + 2 b^2}\right)^{\hspace*{-0.1cm}\frac{N}{2}} \dd b
\Eeq
and
\Beq
    \label{eq:mcvm-dist-odd-d3}
    D^o_3(S) &= \frac{(2L)^2}{(2L + 1)^2} D^e_3(S) + \frac{\bMax^2}{2 (2L + 1)^2} + \int_0^{\bMax} \frac{4b}{(2L + 1)^2} \sum_{i=1}^L \gaussExp{\sqNorm{\vsi}}{2b^2} \dd b \enspace.
\Eeq
The proof is given in \App{app:proof-dist-odd}.
Like for the even case, also the computation of the odd case can be sped up by using the following theorem. 
\begin{Theorem}
\label{theorem-closed-form-odd-d3}\hfill\\
    For a given $\bMax$, the following expression for $D^o_3(S)$ can be obtained
    \Beq
        \label{eq:closed-form-odd-d3}
        D^o_3(S) &= \frac{(2L)^2}{(2L + 1)^2} D^e_3(S) + \frac{\bMax^2}{2 (2L + 1)^2} \,+ \\
                 &\hspace*{0.4cm} \frac{4}{(2L + 1)^2} \sum_{i=1}^L \frac{\bMax^2}{2} \exp\left(-\frac{1}{2}\frac{\sqNorm{\vsi}}{2\bMax^2}\right) + \frac{1}{8} \sqNorm{\vsi} \Ei\left(-\frac{1}{2}\frac{\sqNorm{\vsi}}{2\bMax^2}\right) \enspace,
    \Eeq
    where $\Ei(\cdot)$ denotes the exponential integral.
    \begin{proof}
        The proof is given in \App{app:proof-closed-form-odd-d3}.
    \end{proof}
\end{Theorem}

The extra terms in $D^o_2(S)$ and $D^o_3(S)$, compared to the even case, reflect the influence of the additional sample, placed at the state space origin, on the distance between the Dirac mixture and the standard normal distribution.
The result is that the point mass of the additional sample will cause the other samples to have a slightly larger spread compared to a sample set without the additional sample at the state space origin.
Concerning the above mentioned numerical stability, we also give a proof for the boundedness of both distances $D^e(S)$ and $D^o(S)$ in \App{app:proof-boundedness}.

\subsection{Gradients of the Distance Measures}
\label{sec:distance-measure-gradients}

\newcommand{\sid}{s^{(d)}_i}
\newcommand{\sjd}{s^{(d)}_j}

In order to optimize the parameters $S$ of a given Dirac mixture, we chose to apply a gradient-based iterative optimization procedure.
This requires the partial derivatives of the two distance measures $D^e(S)$ and $D^o(S)$ with respect to the set of parameters $S$.
For the even case, the partial derivatives are
\Beq
    \frac{\partial D^e(S)}{\partial \sid} = -2 \frac{\partial D^e_2(S)}{\partial \sid} + \frac{\partial D^e_3(S)}{\partial \sid} \quad \forall d \in \{ 1, \ldots, N \} \enspace,
\Eeq
with its two terms
\Beq
    \frac{\partial D^e_2(S)}{\partial \sid} = -\frac{\sid}{2L} \int_0^{\bMax} &\frac{2b}{(1 + 2b^2)} \left(\frac{2 b^2}{1 + 2 b^2}\right)^{\hspace*{-0.1cm}\frac{N}{2}} \cdot \gaussExp{\sqNorm{\vsi}}{(1 + 2b^2)} \dd b \enspace,
\Eeq
and
\Beq
    \frac{\partial D^e_3(S)}{\partial \sid} &= -\frac{2}{(2L)^2} \int_0^{\bMax} \frac{1}{b} \cdot \sum_{j=1}^L (\sid - \sjd) \gaussExp{\sqNorm{\vsi - \vsj}}{2b^2} \,+ \\
    &\hspace*{4.1cm} (\sid + \sjd) \gaussExp{\sqNorm{\vsi + \vsj}}{2b^2} \dd b \enspace.
\Eeq
Analogous to $D^e_3(S)$, the following theorem can be used for the computation of its partial derivatives.
\begin{Theorem}
\label{theorem-closed-form-even-deriv-d3}\hfill\\
    For a given $\bMax$, the following expression for $\frac{\partial D^e_3(S)}{\partial \sid}$ can be obtained
    \Beq
        \label{eq:closed-form-even-deriv-d3}
        \frac{\partial D^e_3(S)}{\partial \sid} &= \frac{1}{(2L)^2} \sum_{j=1}^L (\sid - \sjd) \Ei\left(-\frac{1}{2}\frac{\sqNorm{\vsi - \vsj}}{2\bMax^2}\right) + (\sid + \sjd) \Ei\left(-\frac{1}{2}\frac{\sqNorm{\vsi + \vsj}}{2\bMax^2}\right) \enspace,
    \Eeq
    where $\Ei(\cdot)$ denotes the exponential integral.
    \begin{proof}
        The proof is given in \App{app:proof-closed-form-even-deriv-d3}.
    \end{proof}
\end{Theorem}

As with the distance $D^o(S)$ itself, its partial derivatives
\Beq
    \frac{\partial D^o(S)}{\partial \sid} = -2 \frac{\partial D^o_2(S)}{\partial \sid} + \frac{\partial D^o_3(S)}{\partial \sid} \quad \forall d \in \{ 1, \ldots, N \}
\Eeq
can be obtained in terms of the even case plus additional terms according to
\Beq
    \frac{\partial D^o_2(S)}{\partial \sid} = \frac{2L}{2L + 1} \frac{\partial D^e_2(S)}{\partial \sid}
\Eeq
and
\Beq
    \frac{\partial D^o_3(S)}{\partial \sid} &= \frac{(2L)^2}{(2L + 1)^2} \frac{\partial D^e_3(S)}{\partial \sid} - \frac{2\sid}{(2L + 1)^2} \int_0^{\bMax} \frac{1}{b} \gaussExp{\sqNorm{\vsi}}{2b^2} \dd b \enspace.
\Eeq
To ease the computation of the partial derivatives of $D^o_3(S)$, the next theorem can be used.
\begin{Theorem}
\label{theorem-closed-form-odd-deriv-d3}\hfill\\
    For a given $\bMax$, the following expression for $\frac{\partial D^o_3(S)}{\partial \sid}$ can be obtained
    \Beq
        \label{eq:closed-form-odd-deriv-d3}
        \frac{\partial D^o_3(S)}{\partial \sid} &= \frac{(2L)^2}{(2L + 1)^2} \frac{\partial D^e_3(S)}{\partial \sid} + \frac{\sid}{(2L + 1)^2} \Ei\left(-\frac{1}{2}\frac{\sqNorm{\vsi}}{2\bMax^2}\right) \enspace,
    \Eeq
    where $\Ei(\cdot)$ denotes the exponential integral.
    \begin{proof}
        The proof is given in \App{app:proof-closed-form-odd-deriv-d3}.
    \end{proof}
\end{Theorem}

\begin{figure}
    \centering
    \includegraphics[width=0.5\textwidth]{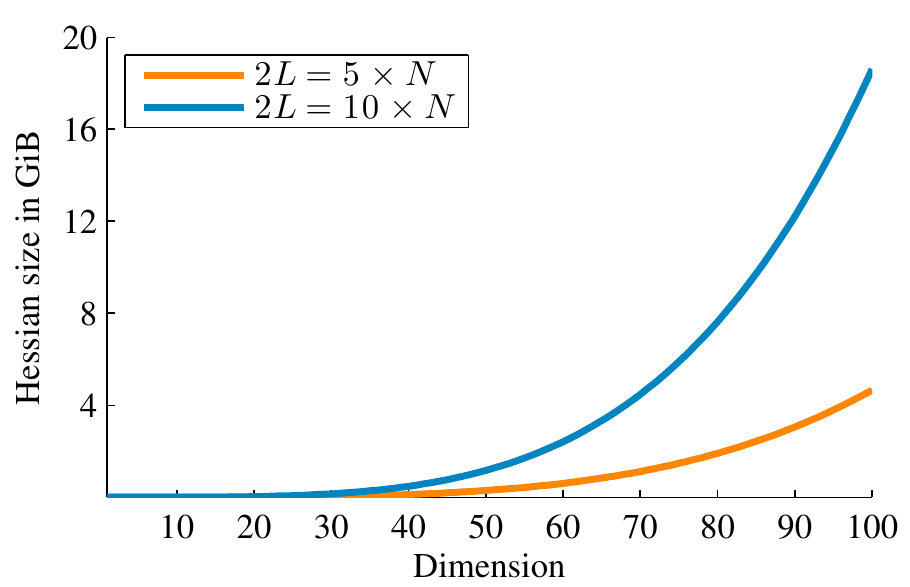}
    \caption{Size of the Hessian matrix for different dimensions and number of samples.}
    \label{fig:hessian-size}
\end{figure}

\subsection{The \sskf with Symmetric Samples}
\label{sec:sym-s2kf}

After defining the distance measures $D^e(S)$ and $D^o(S)$, including their partial derivatives, we can compute a Dirac mixture approximation of a standard normal distribution comprising an arbitrary number of optimally placed point symmetric samples.

To achieve this, we utilize the low memory BFGS quasi-Newton optimization (L-BFGS) \cite{jorge_nocedal_numerical_2006}.
The low memory variant is essential here, as it avoids the explicit computation and storage of the Hessian matrix.
The set of Dirac mixture parameters $S$ encompasses $L \times N$ single parameters to be optimized.
Hence, the Hessian matrix of $D^e(S)$ or $D^o(S)$ would contain $(L \times N)^2$ entries.
When now assuming only a linear increase in the number of samples for an increasing dimension $N$, that is, $2L = C \cdot N$, with a linear factor $C$, the size of the Hessian grows with $\mathcal{O}(N^4)$.
This problem is illustrated in \Fig{fig:hessian-size} for two different linear factors ($5$ and $10$).
It can be seen that approximating a 100-dimensional  standard normal distribution with a thousand samples would require a Hessian of $\approx 20$~gigabytes, and already a Hessian of over $4$~gigabytes in case of only $500$~samples.
Consequently, using the Hessian directly in the optimization is intractable.

The computation of the point symmetric samples works as follows.
\begin{enumerate}
    \item Choose the desired number of samples~$M$ to approximate the $N$-dimensional standard normal distribution.
    \item Depending on the number of samples, the even or odd distance measure is selected.
    \item A proper maximum kernel width $\bMax$  has to be selected.
Generally speaking, the larger the dimension $N$ is the larger $\bMax$ has to be in order to consider all sample positions during the optimization, and thus, to get a meaningful approximation.
Empirically, we have found that a value of $70$ is large enough for up to $N \leq 1000$ dimensions.
    \item An initial set of Dirac mixture parameters is obtained by drawing $L$ samples randomly from an $N$-dimensional standard normal distribution.
    \item The L-BFGS procedure optimizes the Dirac mixture parameters such that the distance measure is minimized, i.e., it moves the initial samples in the state space to approximate the standard normal distribution in an optimal way.
The Dirac mixture parameters resulting from the L-BFGS procedure are denoted as $\{ \vzi \}_{i=1}^{L}$.
    \item The Dirac mixture approximation given by the set of parameters $\{ \vzi \}_{i=1}^{L}$ finally undergoes a Mahalanobis transformation so that the transformed Dirac mixture captures the identity covariance matrix of the standard normal distribution as much as possible.
This is necessary, as the proposed distance measures do not explicitly consider the covariance matrix as a constraint.
The transformation is done by computing the sample covariance matrix
\Beq
    \Cz = \frac{2}{M} \sum_{i=1}^L \vzi \cdot \vzi\T \enspace,
\Eeq
and scaling each sample according to
\Beq
    \vsi = \sqrt{(\Cz)^{-1}} \cdot \vzi \quad \forall i \in \{ 1, \ldots, L \} \enspace,
\Eeq
where $\sqrt{(\Cz)^{-1}}$ denotes the Cholesky decomposition of the inverse sample covariance matrix.
The Dirac mixture defined by the parameters $\{ \vsi \}_{i=1}^{L}$ is the final approximation of the standard normal distribution.
\end{enumerate}
Experimentally, we have found that in situations where the covariance matrix was added as an explicit constraint to the optimization procedure, the sample covariance matrix of the resulting Dirac mixture was less accurate compared to the proposed Mahalanobis approach.
Moreover, the constraint made the optimization procedure much more time-consuming.
Consequently, we dropped this approach in favor of the Mahalanobis transformation.

\begin{figure*}
    \centering
    \begin{subfigure}[b]{0.32\textwidth}
        \centering
        \includegraphics[width=\textwidth]{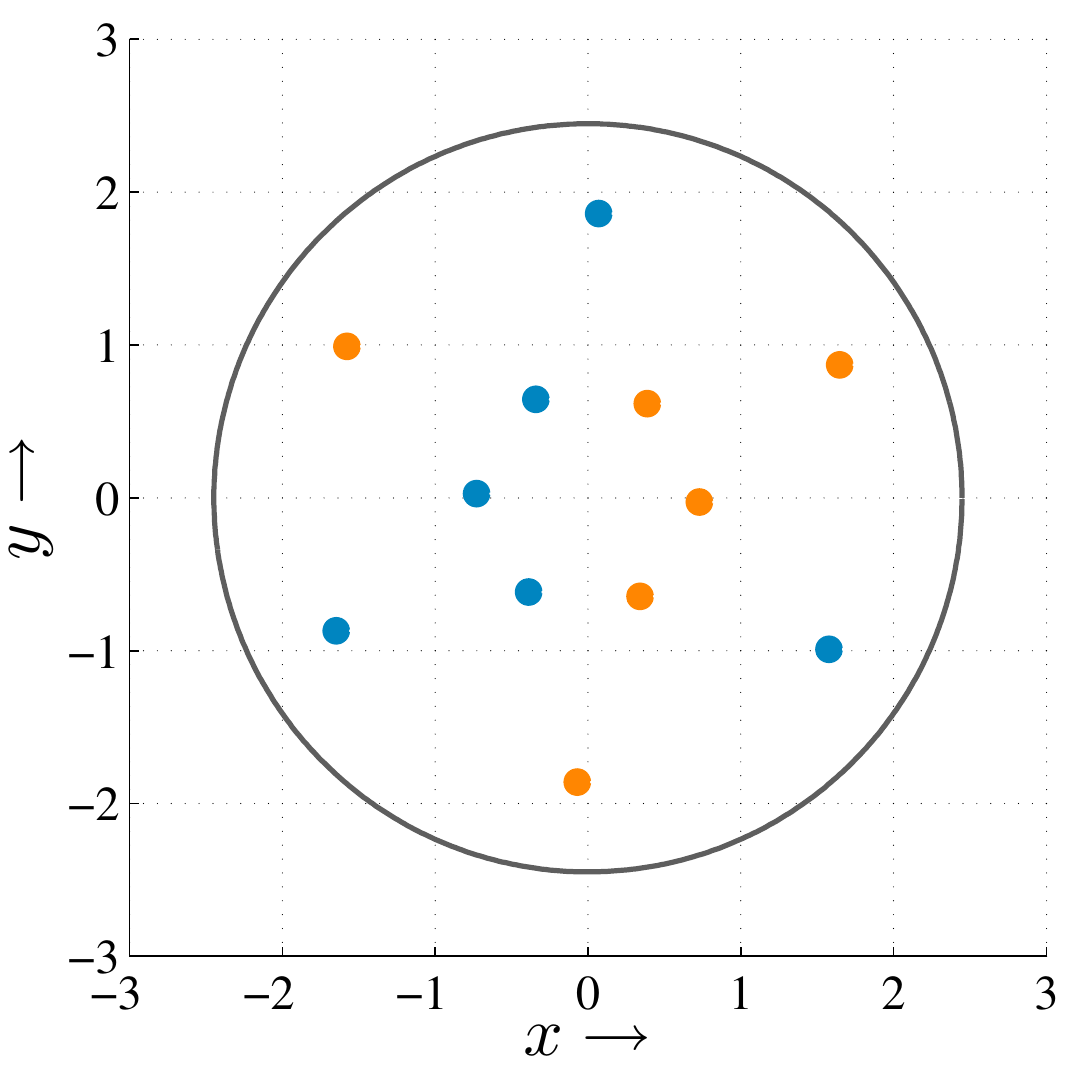}
        \caption{Symmetric approach with 12 samples.}
        \label{fig:s2kf-samplings-sym-12}
    \end{subfigure}
    \hfill
    \begin{subfigure}[b]{0.32\textwidth}
        \centering
        \includegraphics[width=\textwidth]{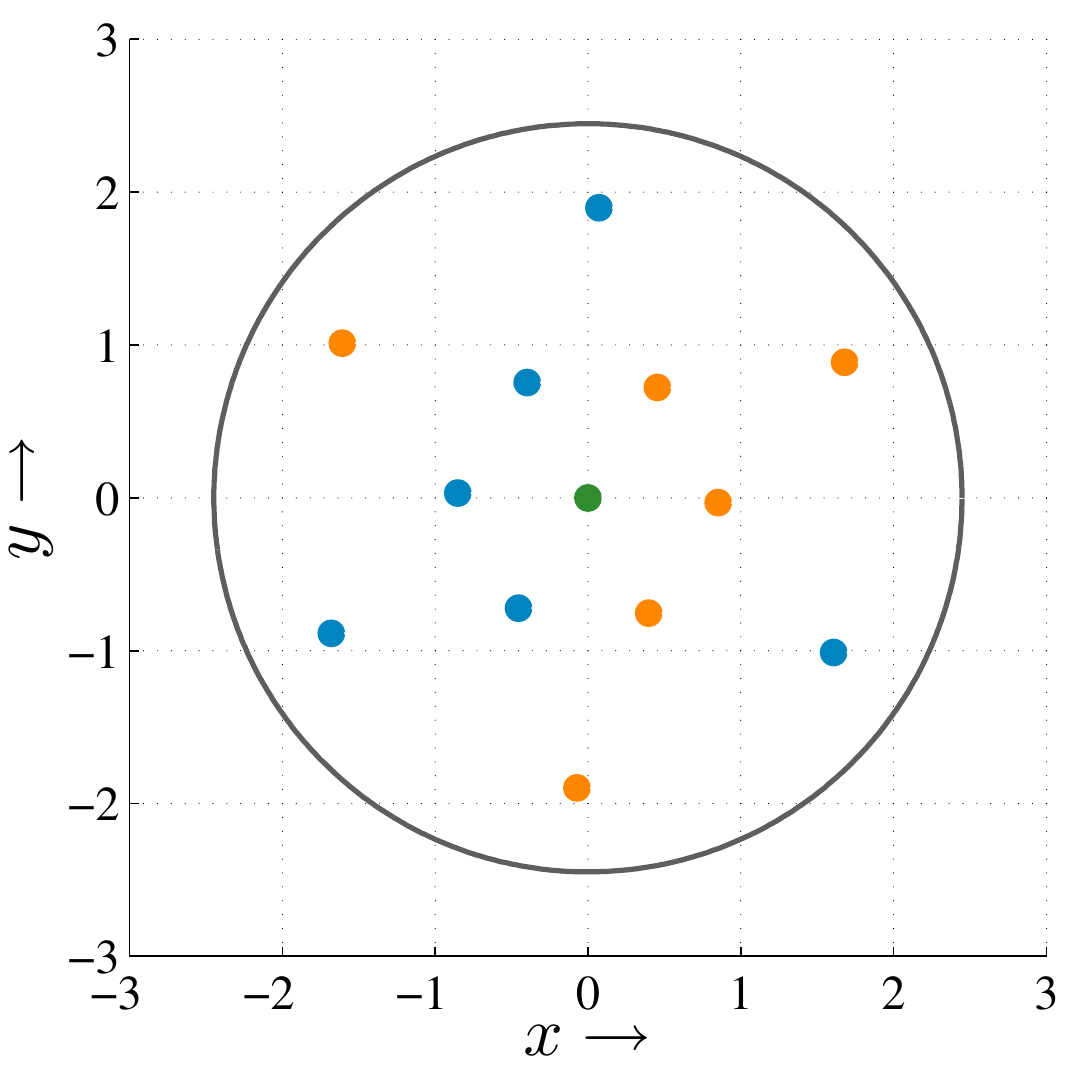}
        \caption{Symmetric approach with 13 samples.}
        \label{fig:s2kf-samplings-sym-13}
    \end{subfigure}
    \hfill
    \begin{subfigure}[b]{0.32\textwidth}
        \centering
        \includegraphics[width=\textwidth]{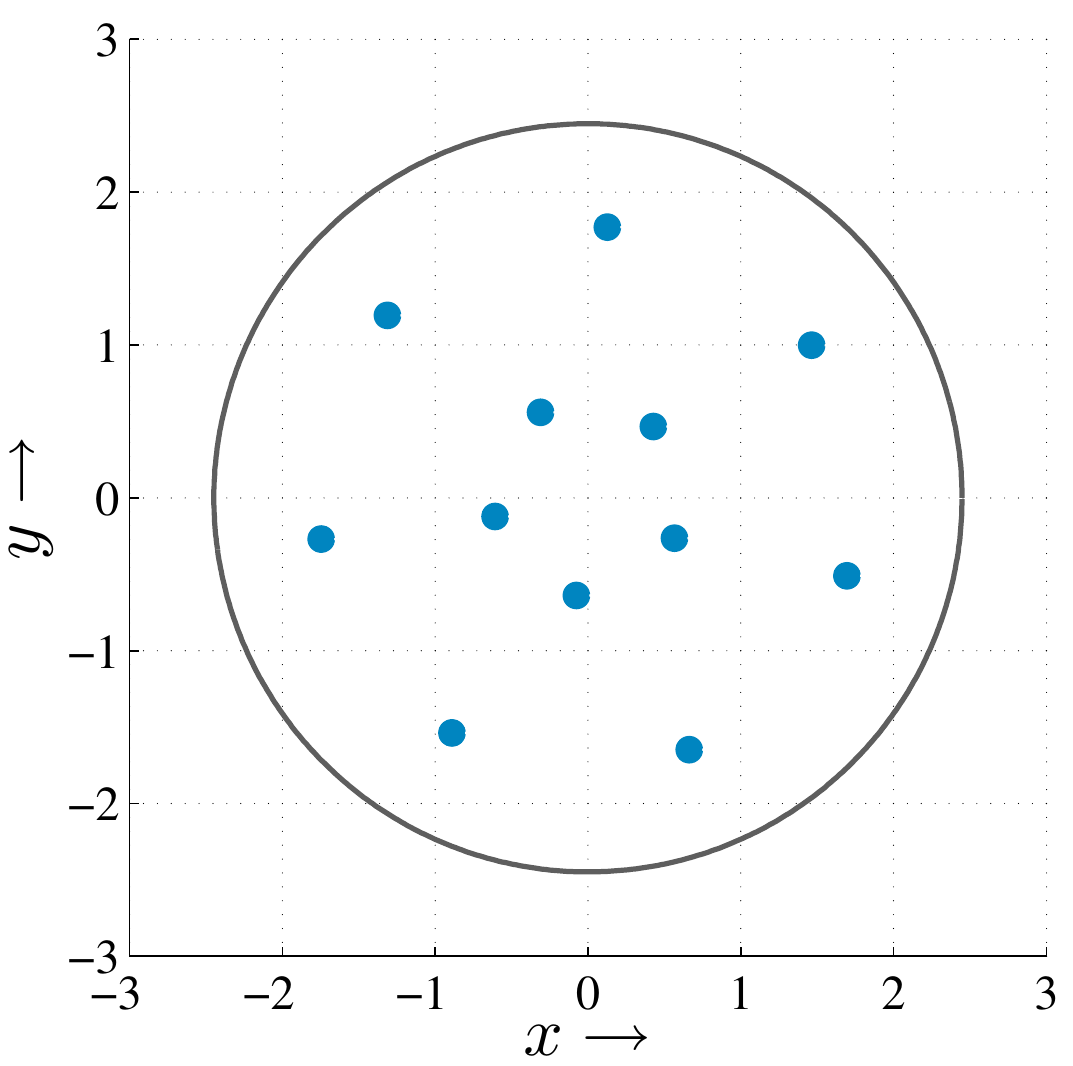}
        \caption{Asymmetric approach with 12 samples.}
        \label{fig:s2kf-samplings-asym-12}
    \end{subfigure}
    \caption{Different LCD-based approximations of a two-dimensional standard normal distribution with samples $\vsi$ (blue), point symmetric counterparts $-\vsi$ (orange), fixed sample at the state space origin in the odd case (green), and $95\%$ confidence interval of the standard normal distribution (gray).}
    \label{fig:s2kf-samplings}
\end{figure*}

The results of different LCD-based approximations of a two-dimensional standard normal distribution are depicted in \Fig{fig:s2kf-samplings}.
On the one hand, Figures \ref{fig:s2kf-samplings-sym-12} and \ref{fig:s2kf-samplings-sym-13} show approximations using the new symmetric sampling scheme comprising $12$ and $13$ samples, respectively.
The point symmetric arrangement around the state space origin can be clearly seen.
Note also the subtle difference in the sample spread of the samples near the state space origin between \Fig{fig:s2kf-samplings-sym-12} and \Fig{fig:s2kf-samplings-sym-13}.
This is caused by the additional point mass from the fixed sample at the state space origin.
On the other hand, \Fig{fig:s2kf-samplings-asym-12} shows an approximation based on the classical asymmetric sampling scheme also comprising $12$ samples.
Here, the optimization procedure can position all samples individually, and hence, the samples are not necessarily arranged in a point symmetric way like in the depicted case.

Using the above described optimal point symmetric sampling of a standard normal distribution, we obtain a symmetric version of the \sskf.
Furthermore, to avoid a re-computation on every program start, we store any computed Dirac mixture approximation of a standard normal distribution persistent in the file system for later reuse.
This mechanism is called the \emph{Sample Cache} and was already used by the asymmetric \sskf.

\section{Evaluation}
\label{sec:evaluation}

In this Section, we want to compare the new point symmetric sampling scheme of the \sskf with its asymmetric version and other state-of-the-art LRKFs.
First, we take a closer look at the approximation of higher-order moments of standard normal distributions.
Then, the advantage of using a point symmetric sampling scheme, and hence, the new version of the \sskf, is discussed by means of a simple symmetric measurement equation.
Finally, extended object tracking is performed to compare the recursive state estimation quality of various state-of-the-art LRKFs.

\subsection{Moment Errors of a Standard Normal Distribution}

First, we investigate how well the employed sampling schemes of state-of-the-art LRKFs approximate the moments of a standard normal distribution.
Thus, we are interested in the expectation values
\Beq
    \E{x_1^{n_1} x_2^{n_2} \cdots x_N^{n_N}} = \int_{\IR^N} x_1^{n_1} x_2^{n_2} \cdots x_N^{n_N} \Gauss(\vx\,; \vzero, \mI_N) \dd \vx \enspace,
\Eeq
with
\Beq
    \sum_{i=1}^N n_i = m \;, \quad 0 \leq n_i  \leq m
\Eeq
for different dimensions $N$ and moment orders $m$.
This has the advantage of being independent of a concrete system and measurement model.
Note that for given $N$ and $m$, there exists $N^m$ possible combinations to select the values for $n_i$.
Hence, a moment is characterized by $N^m$ values.

As all state-of-the-art LRKFs employ a point symmetric sampling scheme and capture mean and covariance matrix, we focus on higher-order even moments.
More precisely, we take a look at the 4th, 6th, and 8th moment, i.e., $m \in \{ 4, 6, 8 \}$.
In many practical applications, 3D and 6D Gaussian distributions are of special interest.
For example, the location and orientation in 2D or the position in 3D can be estimated using a three-dimensional system state.
When additionally considering velocities in the 2D case or the orientation in the 3D case, a six-dimensional state is required.
Thus, we chose to study the approximations of standard normal distributions with these two dimensions, i.e., $N \in \{ 3, 6 \}$.

To compare the different LRKF sampling techniques, for each dimension $N$ and moment $m$ we compute a normalized moment error according to
\Beq
    \sqrt{\frac{1}{N^m} \sum_{j=1}^{N^m} (\mathbb{E}^{\operatorfont true}_{j} - \mathbb{E}^{\operatorfont LRKF}_{j})^2} \enspace,
\Eeq
where $\mathbb{E}_{j}$ denotes one of the $N^m$ possible combinations for the $m$th moment, the superscript "${\operatorfont true}$" the true moment value and "${\operatorfont LRKF}$" the LRKF sampling estimate.
Note that the 8th moment of the 6D standard normal distribution is characterized by already over 1.5 million combinations.

\begin{figure*}[ht]
    \centering
    \begin{subfigure}[b]{0.32\textwidth}
        \centering
        \includegraphics[width=\textwidth]{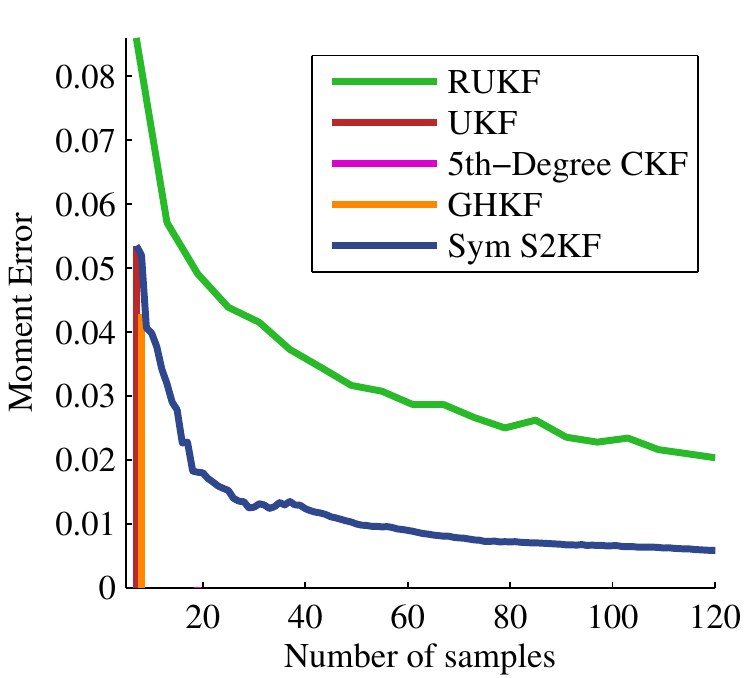}
        \caption{Error 4th moment.}
    \end{subfigure}
    \hfill
    \begin{subfigure}[b]{0.32\textwidth}
        \centering
        \includegraphics[width=\textwidth]{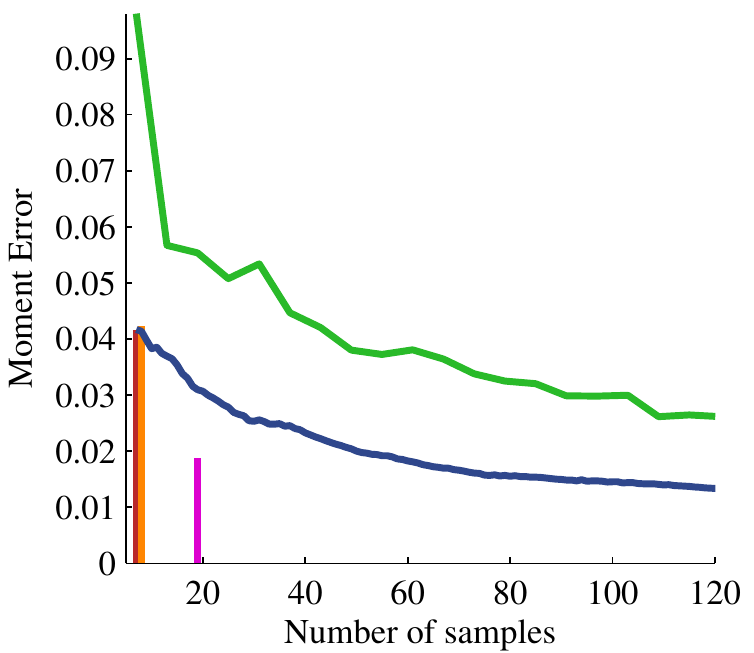}
        \caption{Error 6th moment.}
    \end{subfigure}
    \hfill
    \begin{subfigure}[b]{0.32\textwidth}
        \centering
        \includegraphics[width=\textwidth]{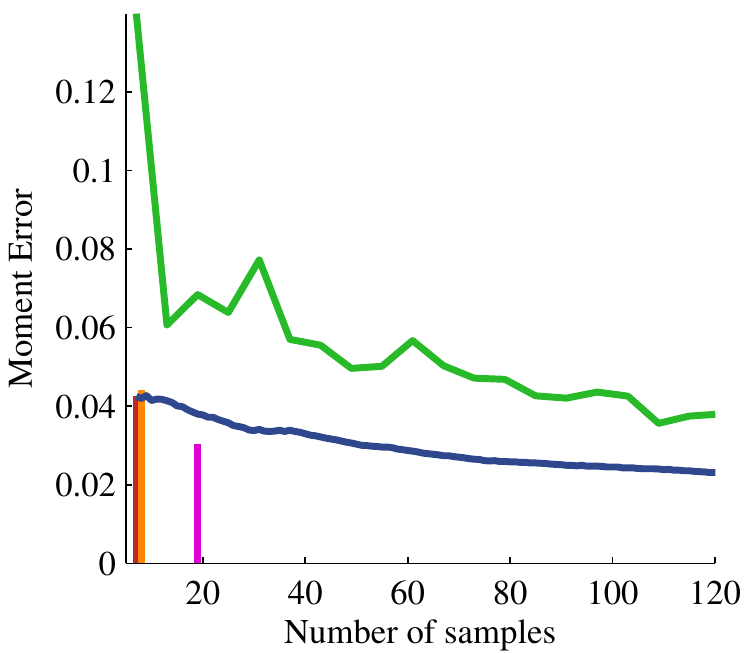}
        \caption{Error 8th moment.}
    \end{subfigure}
    \caption{Moment errors of a 3D standard normal distribution.}
    \label{fig:moment-errors-3d}
\end{figure*}

\begin{figure*}[ht]
    \centering
    \begin{subfigure}[b]{0.32\textwidth}
        \centering
        \includegraphics[width=\textwidth]{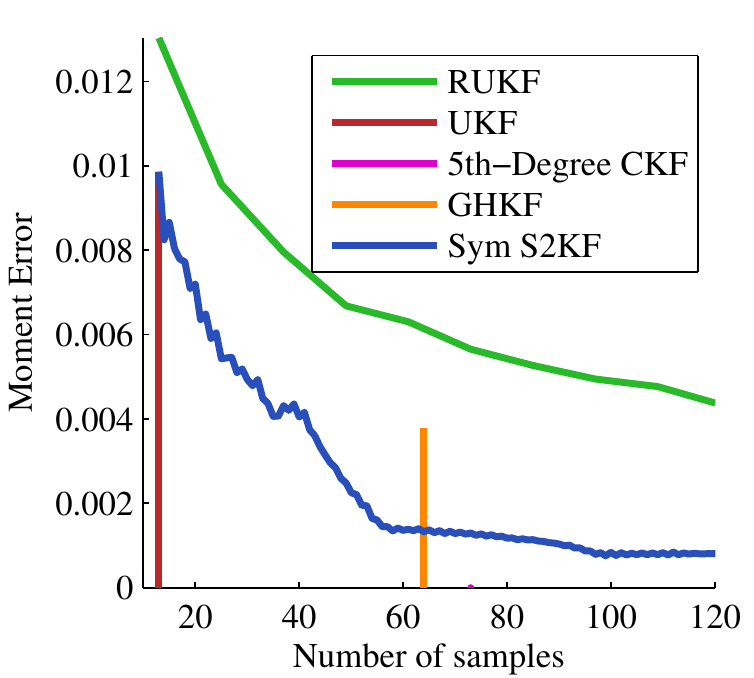}
        \caption{Error 4th moment.}
    \end{subfigure}
    \hfill
    \begin{subfigure}[b]{0.32\textwidth}
        \centering
        \includegraphics[width=\textwidth]{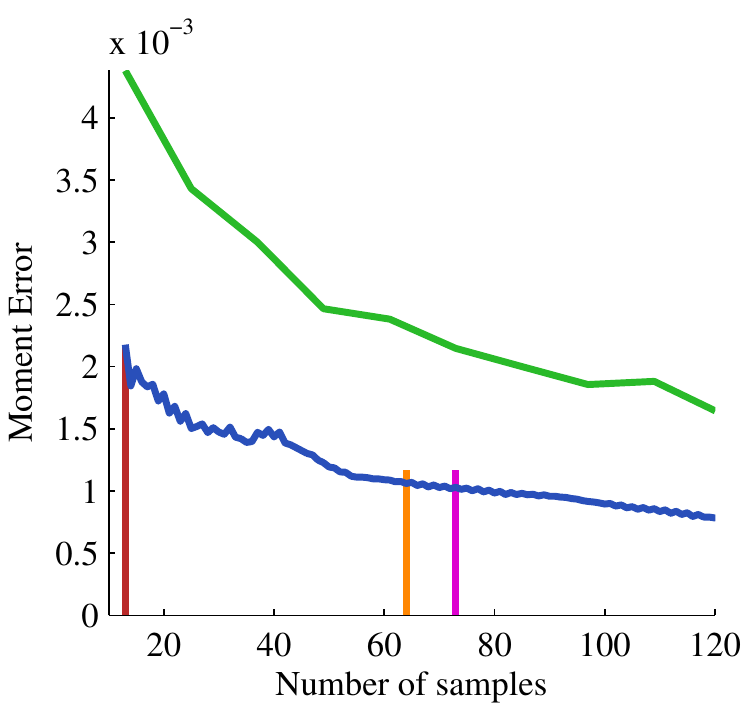}
        \caption{Error 6th moment.}
    \end{subfigure}
    \hfill
    \begin{subfigure}[b]{0.32\textwidth}
        \centering
        \includegraphics[width=\textwidth]{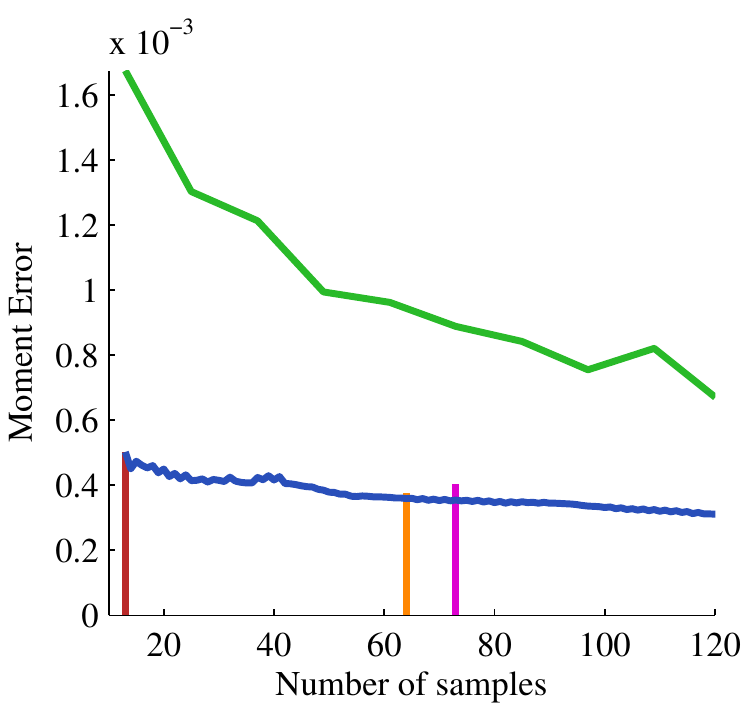}
        \caption{Error 8th moment.}
    \end{subfigure}
    \caption{Moment errors of a 6D standard normal distribution.}
    \label{fig:moment-errors-6d}
\end{figure*}

We compare the new symmetric \sskf, the UKF with equally weighted samples, the RUKF, the \ckf, and the GHKF with two quadrature points.
As the symmetric \sskf and the RUKF do not have unique sample sets, we perform 100 Monte Carlo runs for both filters.
In each run, for both filters new sample sets are generated, and the average moment error over all runs are computed.
Moreover, the \sskf and the RUKF are evaluated with different number of samples.
The results are depicted in Figures \ref{fig:moment-errors-3d} and \ref{fig:moment-errors-6d}.
As the UKF, the \ckf, and the GHKF have a fixed number of samples, they are depicted as a bar at their respective employed number of samples.

The moment errors of the UKF and the \sskf are identical for the case when both filters use the same number of samples.
This is due to the fact that both sample sets are equally weighted and the \sskf places its samples like the UKF (except for the rotation) as this minimize the utilized distance measure.
The RUKF, however, scales the utilized UKF sample sets randomly. Consequently, its sample set is not necessarily equally weighted like for the UKF and the \sskf, and hence, their moment errors differ.
Considering all moments, the \sskf delivers always smaller errors than the RUKF and for nearly all moments smaller errors than the GHKF (for the same number of samples).
The sampling of the \ckf is the only one that matches the 4th moment exactly.
This is based on the fact that the spherical-radial rule of the \ckf has a 5th-degree accuracy~\cite{bin_jia_high-degree_2013}.

\subsection{Symmetric Measurement Equations}

To illustrate the advantages of using a symmetric sampling scheme, we consider the two-dimensional system state
\Beq
    \vx = [a,\;b]\T
\Eeq
combined with the scalar and symmetric measurement equation
\Beq
    \label{eq:sym-meas-model}
    y = h(\vx, v) = \sqrt{a^2 + b^2} + v \enspace,
\Eeq
where $v$ is zero-mean Gaussian noise with variance $\sigma^2 = 0.01$.
Hence, we measure a noisy distance from the system state~$\vx$ to the state space origin.
Such a symmetric measurement equation arises for example in \cite{marcus_baum_kernel-sme_2013, christof_chlebek_bayesian_2014}.

\begin{figure}[ht]
    \centering
    \begin{tikzpicture}
        \node at (0,0) {\includegraphics[width=0.47\textwidth]{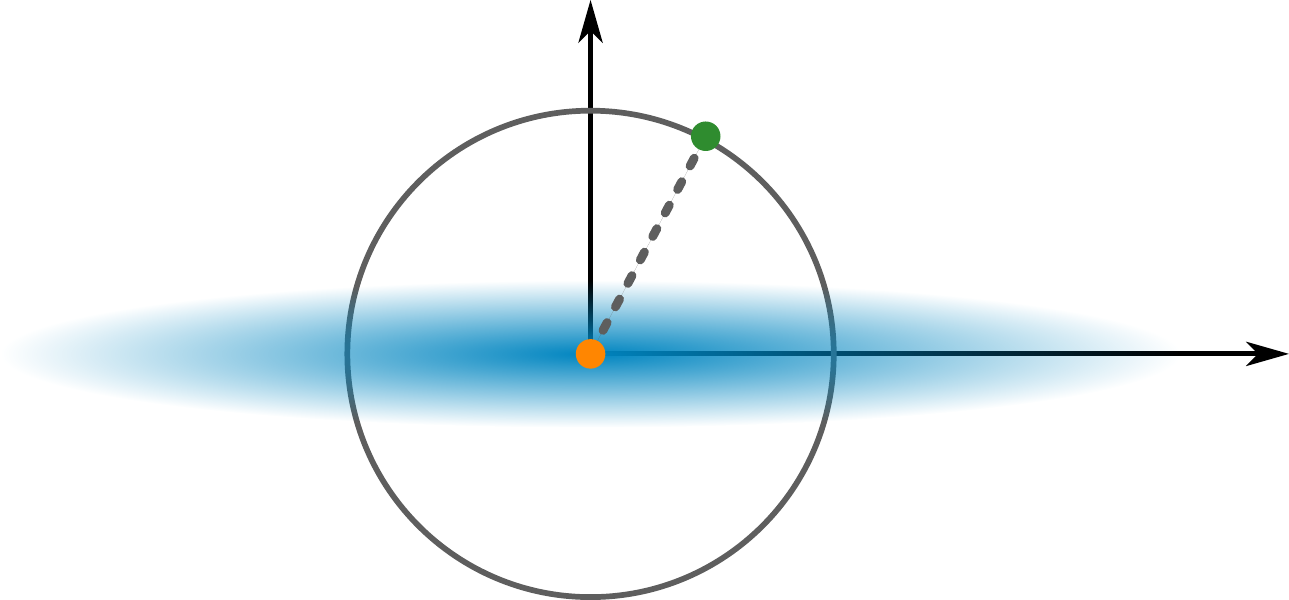}};
        
        \node at (3.7cm,-0.6cm) {$a$};
        \node at (-0.6cm,1.6cm) {$b$};
        
        \node at (-0.0cm,0.7cm) {$y$};
        \node at (-0.3cm,-0.8cm) {$\evx^p$};
        \node at (0.9cm,1.25cm) {$\vx_{true}$};
    \end{tikzpicture}  
    \caption{Symmetric measurement model with prior mean (orange), prior uncertainty (blue), true system state (green).}
    \label{fig:sym-meas-model}
\end{figure}
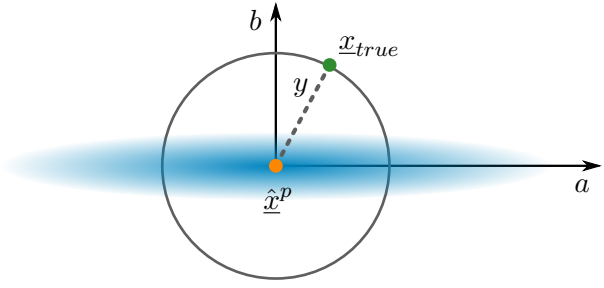

We assume that the true system state is
\Beq
    \vx_{true} = [1,\; 2]\T \enspace,
\Eeq
and our goal is to estimate it using a Nonlinear Kalman Filter initialized with mean and covariance matrix
\Beq
    \evx^p = [0,\; 0]\T,\; \mC^p = \diag(4, 0.5) \enspace.
\Eeq
The setup is illustrated in \Fig{fig:sym-meas-model}.
From the the estimator's perspective, the received measurement $\meas$ could stem from any state located on the gray circle around the prior mean, not only $\vx_{true}$.
Hence, a Nonlinear Kalman Filter cannot gain any new information about the hidden system state from the measurement $\meas$.
This situation is reflected in a zero cross-covariance matrix of state and measurement $\Cxy$ in \Eq{eq:linear-meas-update-mean} and \Eq{eq:linear-meas-update-cov}.
Consequently, the posterior state estimate (mean and covariance matrix) equals the prior, no matter what prior uncertainty we have.

Now, we try to reproduce this result when using LRKFs.
More precisely, we compare the asymmetric \sskf, its new symmetric version (both using $11$ samples), and the UKF.
We perform $R = 100$ Monte Carlo runs.
In each run, we reset the initial state estimate, and simulate a noisy measurement $\meas$ to perform one measurement update.
Moreover, both \sskf variants compute a new set of samples approximating a standard normal distribution in every Monte Carlo run.
We compute the Root Mean Square Error (RMSE) for the posterior mean
\Beq
    \sqrt{\frac{1}{R} \sum_{r=1}^R \norm{\evx^e_{r} - \evx^p}^2} \enspace,
\Eeq
where $\evx^e_{r}$ denotes the estimated posterior mean of run $r$.
Additionally, we compute the RMSE of the posterior covariance matrix
\Beq
    \sqrt{\frac{1}{R} \sum_{r=1}^R \|\mC^e_{r} - \mC^p\|^2} \enspace,
\Eeq
where $\mC^e_{r}$ denotes estimated posterior covariance matrix of run $r$ and $\|\cdot\|$ the Frobenius norm.

\begin{figure}
    \centering
    \includegraphics[width=0.6\textwidth]{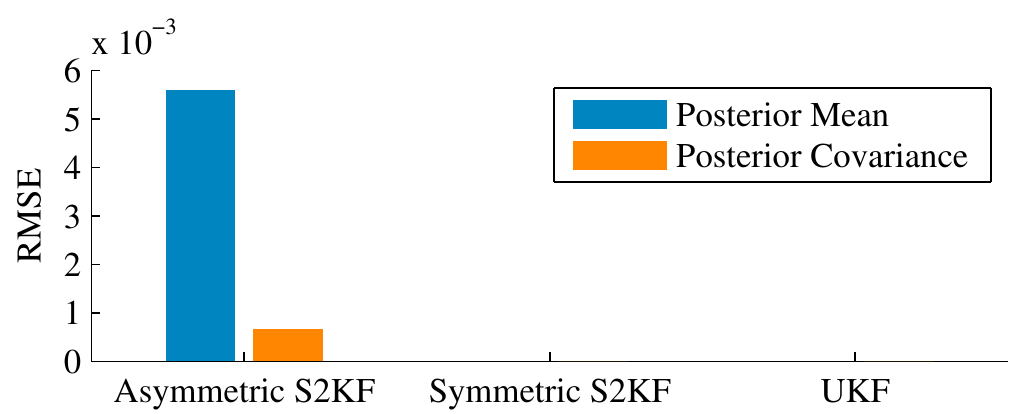}
    \caption{Estimation errors for symmetric measurement model.}
    \label{fig:sym-meas-model-rmse}
\end{figure}

The results of the evaluation are depicted in \Fig{fig:sym-meas-model-rmse}.
It can be seen that the asymmetric \sskf is the only filter with a small error.
This can be explained with its asymmetric sampling scheme, as it makes a system state on the circle around the prior mean more likely than others.
Hence, it introduces (theoretically non-existent) correlations between the measurement and the system state, i.e., a non-zero cross-covariance matrix.
As a consequence, the asymmetric \sskf updates its state estimate mistakenly (albeit not that much).
The other estimators do not have such a problem due to their symmetric sampling.
So even such a simple scenario demonstrates the advantages of the new point symmetric sampling scheme of the \sskf.

\subsection{Extended Object Tracking}

Now, we consider estimating the pose and extent of a cylinder in 3D based on a Random Hypersurface Model (RHM) \cite{florian_faion_tracking_2012, marcus_baum_extended_2014}.
The system state is composed of position $\vck~=~[c^x_k,\, c^y_k,\, c^z_k]\T$ and velocity $\vnuk~=~[\nu^x_k,\, \nu^y_k,\, \nu^z_k]\T$, rotation angles $\vec{\phi}_k~=~[\phi^x_k,\, \phi^y_k]\T$ and their velocities $\vec{\omega}_k~=~[\omega^x_k,\, \omega^y_k]\T$, as well as the cylinder radius $r_k$ and length $l_k$ according to
\Beq
    \vxk = [\vck\T, \vnuk\T, \vec{\phi}_k\T, \vec{\omega}_k\T, r_k, l_k]\T \enspace.
\Eeq

The temporal evolution of the cylinder is modeled with a constant velocity model
\Beq
    \vxk = \mA \vxkk + \vw \enspace,
\Eeq
with system matrix
\Beq
    \mA = \Bmat \mI_3  & \mI_3  & \mzero & \mzero & \mzero \\
                \mzero & \mI_3  & \mzero & \mzero & \mzero \\
                \mzero & \mzero & \mI_2  & \mI_2  & \mzero \\
                \mzero & \mzero & \mzero & \mI_2  & \mzero \\
                \mzero & \mzero & \mzero & \mzero & \mI_2  \Emat
\Eeq
and zero-mean Gaussian white noise $\vw$ with covariance matrix
\Beq
    \Cw = \diag(10^{-6} \mI_3, 
                10^{-4} \mI_3, 
                10^{-10} \mI_2,
                10^{-5} \mI_2,
                10^{-4} \mI_2) \enspace.
\Eeq
This linear model allows to compute the prediction step analytically for all LRKFs.

A measurement is a noisy point
\Beq
    \vmeask = [\meas^x_k,\, \meas^y_k,\, \meas^z_k]\T
\Eeq
from the cylinder's surface.
It is related to the system state by means of the implicit nonlinear measurement equation
\Beq
    \label{eq:cylinder-meas-model}
    \vzero &= \vh(\vxk, \vmeask, \vv, s) = \Bmat (m^x_k)^2 + (m^y_k)^2 - r_k^2 \\
                                                m^z_k - s \cdot l_k      \\
                                               (m^z_k - s \cdot l_k)^2   \Emat \enspace,
\Eeq
where
\Beq
    \vmk = (\mR(\phi^y_k) \cdot \mR(\phi^x_k))^{-1}(\vmeask - \vv - \vck) \enspace,
\Eeq
and zero-mean Gaussian white noise $\vv$ with covariance matrix $\Cv = 0.01 \cdot \mI_3$ and multiplicative white noise $s~\hspace*{-0.05cm}\sim~\hspace*{-0.05cm}\Uniform(-0.5, 0.5)$\;\footnote{As LRKFs can only sample Gaussian distributions, the uniform distribution will be approximated as a Gaussian using moment matching.}.
Furthermore, $\mR(\cdot)$ denotes a 3D rotation matrix around the respective axis.
It is important to note that the measurement equation itself depends on the received measurement $\vmeask$, and the estimator only takes the so-called pseudo measurement $\vzero$ as input.
The reason for this is that the proposed measurement model tries to minimize the Euclidean distance between the received measurements $\vmeask$ and the cylinder's surface, and thus, generates measurements of value zero in the optimal case.
Note also that the quadratic term in the last row of \Eq{eq:cylinder-meas-model} is necessary when dealing with multiplicative noise in combination with Kalman Filters \cite{marcus_baum_modeling_2012, florian_faion_tracking_2012}.

At each time step, we receive a set of $20$ measurements
\Beq
    \meassetk = \{ \vmeask^{(1)}, \ldots, \vmeask^{(20)} \} \enspace.
\Eeq
As the order of processing measurements affects the filtered state estimate, we do not process measurements sequentially.
More precisely, we process all measurements \emph{at once}, that is, in a single measurement update, by stacking the measurements into a large measurement vector according to
\Beq
    \Bmat \vzero \\[0.1cm]
          \vdots \\[0.1cm]
          \vzero \Emat =
    \Bmat \vh(\vxk, \vmeask^{(1)}, \vv^{(1)}, s^{(1)})    \\
          \vdots                                          \\
          \vh(\vxk, \vmeask^{(20)}, \vv^{(20)}, s^{(20)}) \Emat \enspace.
\Eeq
This, in turn, requires a set of $20 \cdot 4 = 80$ measurement noise variables in total.
Together with the twelve-dimensional system state, a LRKF has to sample a 92-dimensional random vector to perform a measurement update.
The number of samples used by the investigated LRKFs are summarized in \Tab{tab:cylinder-tracking-setup}.
It should be noted that the GHKF \cite{kazufumi_ito_gaussian_2000} is intractable for the considered scenario as it relies on a Cartesian product and would require at least $2^{92}$ samples.

\begin{table}
    \centering
    
\begin{tabular}{l l l r}
    \toprule
    LRKF & \multicolumn{3}{l}{Number of samples} \\
    \midrule
    Fifth-degree CKF & $2 \cdot 92^2 + 1$ & $=$ & $16,929$ \\
    RUKF (with 5 iterations)& $5 \cdot (2 \cdot 92) + 1$ & $=$ & $921$ \\
    RUKF (with 20 iterations)& $20 \cdot (2 \cdot 92) + 1$ & $=$ & $1,841$ \\
    Asymmetric \sskf & \multicolumn{2}{l}{Freely selectable} & $461$ \\
    Asymmetric \sskf & \multicolumn{2}{l}{Freely selectable} & $1,841$ \\
    Symmetric \sskf & \multicolumn{2}{l}{Freely selectable} & $461$ \\
    Symmetric \sskf & \multicolumn{2}{l}{Freely selectable} & $1,841$ \\
    \bottomrule
\end{tabular}

    \caption{LRKF settings for the measurement update.}
    \label{tab:cylinder-tracking-setup}
\end{table}

\begin{figure}
    \centering
    \includegraphics[width=0.6\textwidth]{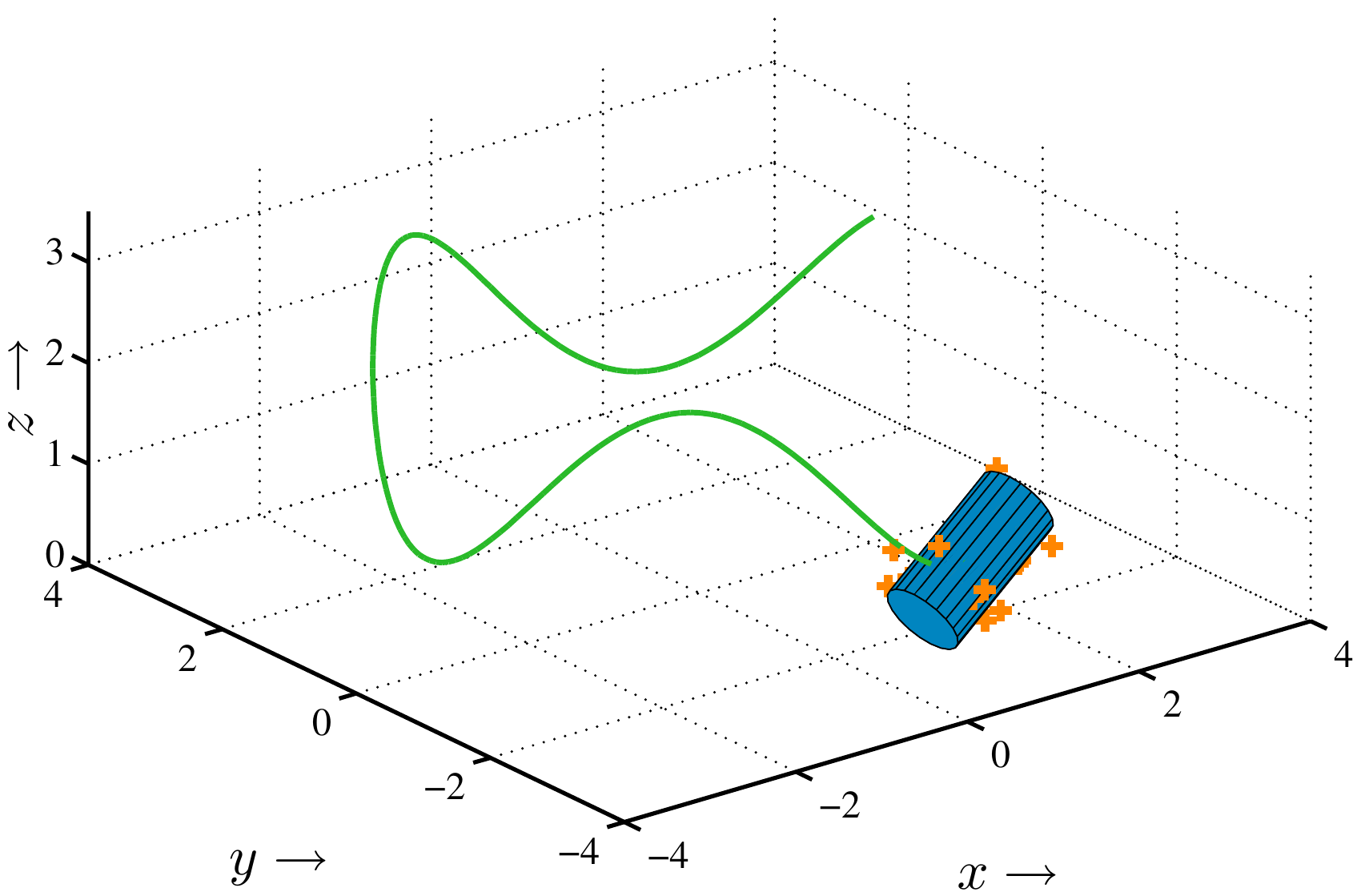}
    \caption{Cylinder state (blue) after $360$ time steps inclusive its trajectory (green line) and $20$ noisy measurements (orange crosses).}
    \label{fig:cylinder-tracking-setup}
\end{figure}

\begin{figure*}
    \centering
    \begin{subfigure}[b]{0.49\textwidth}
        \centering
        \includegraphics[width=\textwidth]{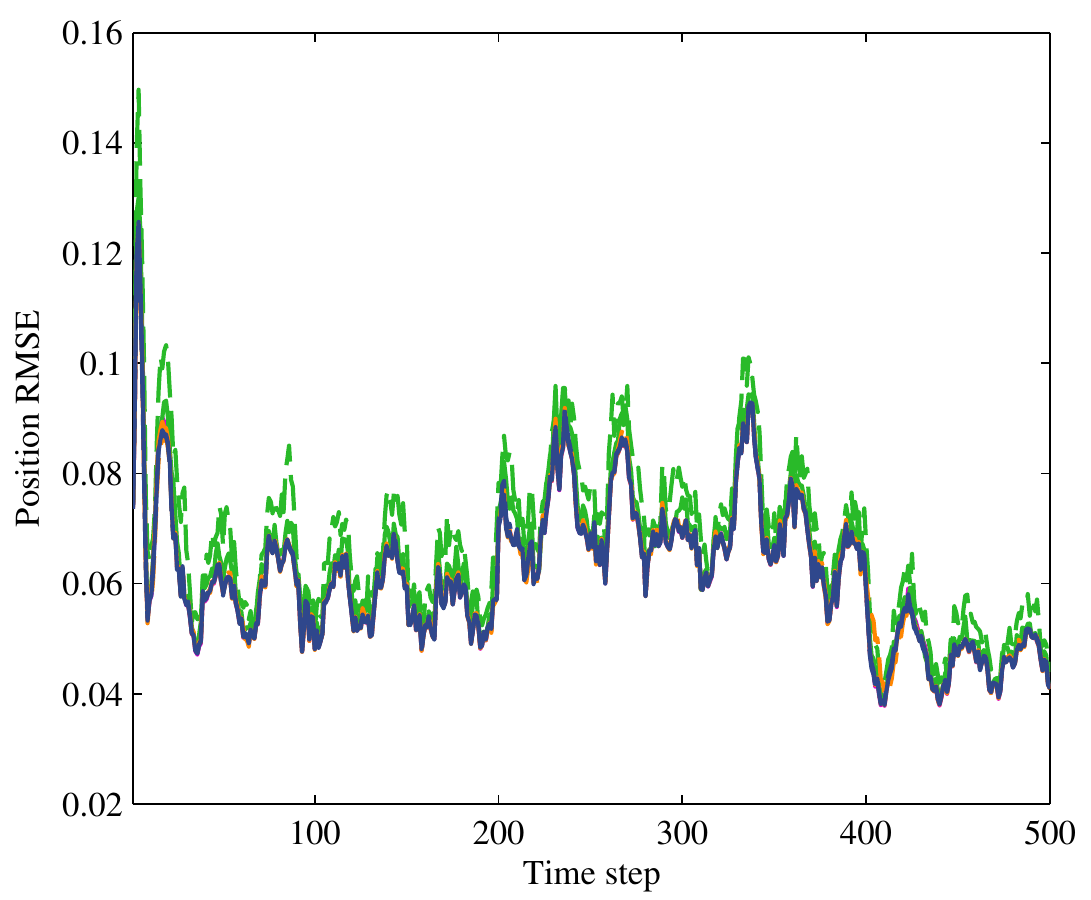}
        \caption{Cylinder position error.}
        \label{fig:cylinder-tracking-position-error}
    \end{subfigure}
    \hfill
    \begin{subfigure}[b]{0.49\textwidth}
        \centering
        \includegraphics[width=\textwidth]{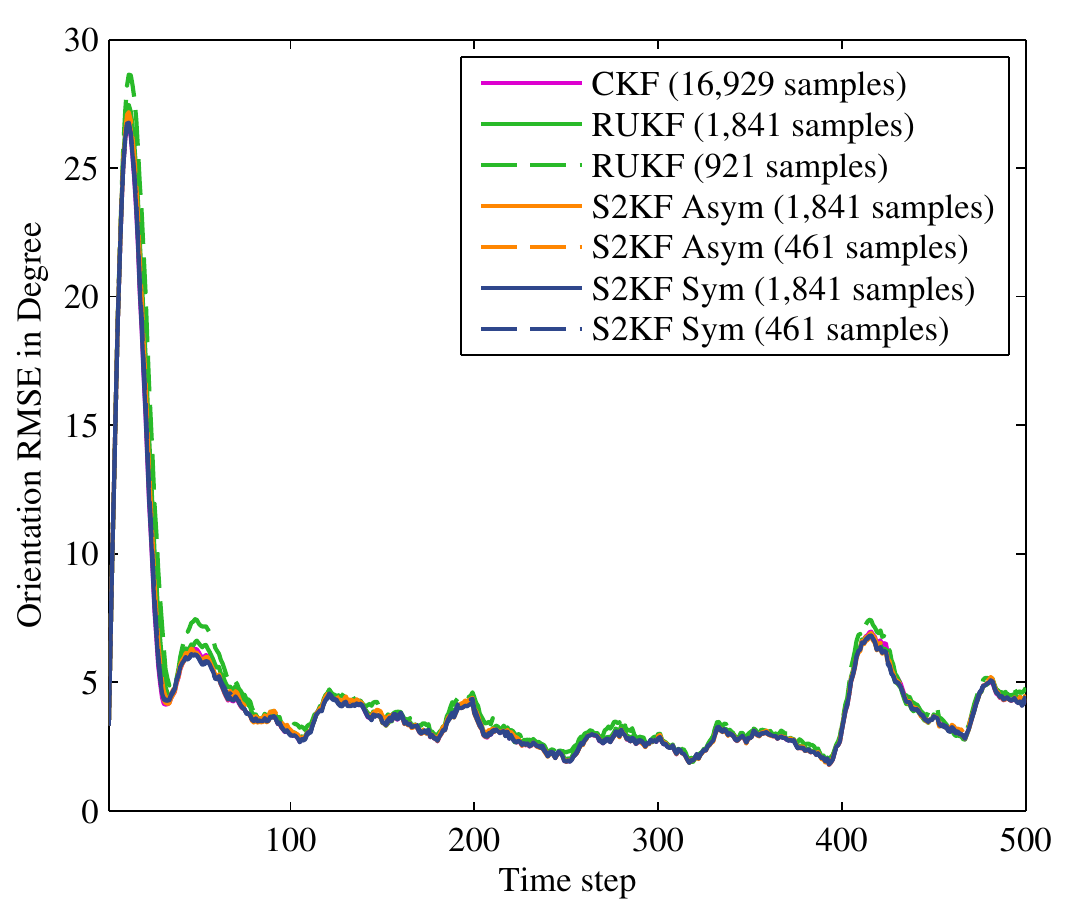}
        \caption{Cylinder orientation error.}
        \label{fig:cylinder-tracking-orientation-error}
    \end{subfigure}
    \begin{subfigure}[b]{0.49\textwidth}
        \centering
        \includegraphics[width=\textwidth]{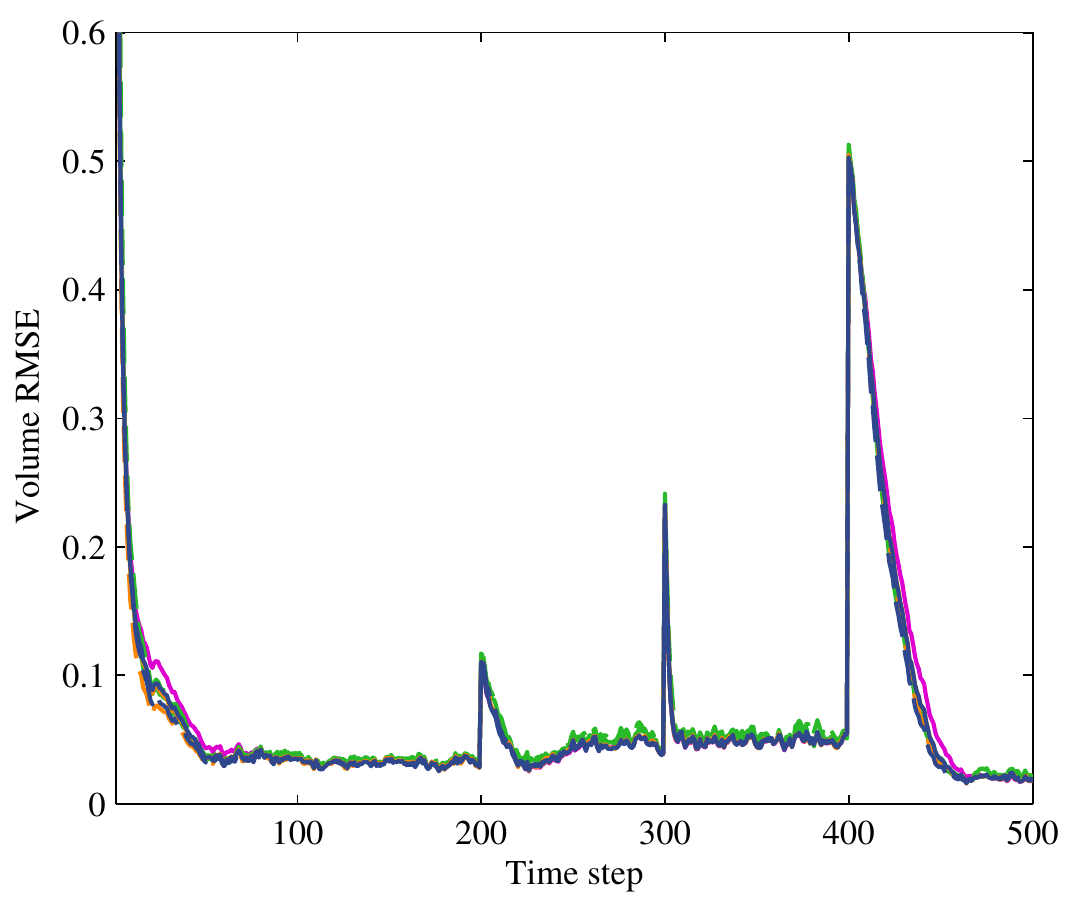}
        \caption{Cylinder volume error.}
        \label{fig:cylinder-tracking-volume-error}
    \end{subfigure}
    \hfill
    \begin{subfigure}[b]{0.49\textwidth}
        \centering
        \includegraphics[width=\textwidth]{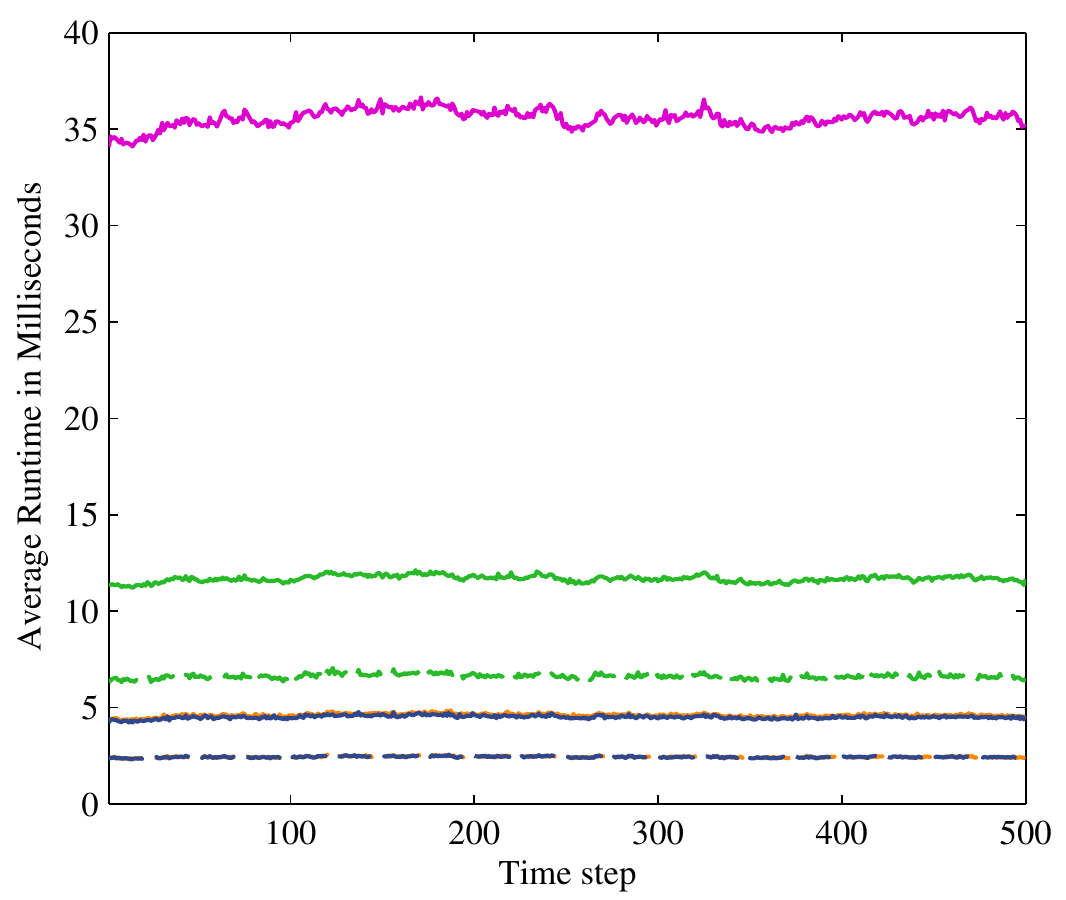}
        \caption{Measurement update runtime.}
        \label{fig:cylinder-tracking-runtime}
    \end{subfigure}
    \caption{Cylinder tracking simulation results.}
    \label{fig:cylinder-tracking-results}
\end{figure*}

We simulate a nonlinear trajectory of a cylinder over 500 time steps including rotations in all its three degrees of freedom as depicted in \Fig{fig:cylinder-tracking-setup}.
Additionally, the initial cylinder's length of $1$ increases to $1.5$ after $200$ time steps, and the initial radius of $0.3$ increases to $0.4$ after further $100$ time steps.
Finally, at time step $400$, the cylinder's length shrinks back to $0.5$.
We perform $100$ Monte Carlo runs.
In reach run, we initialize the estimators with
\Beq
    \evx^e_0 &= [\evc\T, 0, \ldots, 0, 1, 2]\T \\
    \mC^e_0  &= \diag(\mC^c, 10^{-3} \mI_3, 10^{-7} \mI_4, 10^{-2} \mI_2) \enspace,
\Eeq
where $\evc$ denotes the mean and $\mC^c$ the covariance of the first set of measurements $\measset_0$.
For each investigated LRKF, we compute the cylinder position RMSE (\Fig{fig:cylinder-tracking-position-error}), the RMSE of the angle between the true cylinder longitudinal axis and the estimated one (\Fig{fig:cylinder-tracking-orientation-error}), as well as the cylinder volume RMSE (\Fig{fig:cylinder-tracking-volume-error}).
Regarding the cylinder position, the RUKF instances were the filters with the largest errors although they used the same or twice the number of samples of the \sskf variants.
The asymmetric \sskf was a little bit less accurate than the symmetric \sskf and the \ckf.
Same results can be observed for the cylinder orientation error.
For the cylinder volume error, all estimators had noticeable error peaks at time steps $200$, $300$, and $400$.
These can be explained with the abrupt shape changes of the cylinder at the respective time steps.
Furthermore, the \ckf is not as good as in the other estimation quality criteria, and also the asymmetric \sskf is slightly better than the symmetric \sskf in the beginning.

However, when looking at the runtimes of the respective LRKF measurement updates in \Fig{fig:cylinder-tracking-runtime}, the \ckf was the slowest filter due to its large amount of samples.
The runtimes of the asymmetric and the symmetric \sskf were nearly identical as they used the same number of samples.
For the case when the RUKF and the \sskf variants used the same number of samples, the RUKF was slower (11.5~ms compared to 4.5~ms) due to the additional overhead resulting from the creation of several 92-dimensional random orthogonal matrices during each measurement update.
All in all, both \sskf variants were the filters yielding the best compromise between runtime performance and estimation accuracy.
Moreover, this illustrates the advantage of being able to select the number of samples independently of the state/noise dimensions, in contrast to the \ckf.

\section{Conclusions}
\label{sec:conclusions}

In this paper, we introduced a new point symmetric Gaussian sampling scheme for the Smart Sampling Kalman Filter.
This reflects the point symmetry of the Gaussian distribution and allows for matching all odd moments of a standard normal distribution exactly.
The new sampling technique does not only improve the estimation quality of the \sskf, it also speeds up the computation of the Gaussian samples as the number of Dirac mixture parameters to be optimized is reduced by half.

After describing the structure of a sample-based Kalman Filter, we extended the general Dirac mixture to a point symmetric form by distinguishing between an even and an odd number of samples.
Then, we adapted the existing LCD distance measure to these new Dirac mixtures and also gave formulas for their respective gradients.
These are required by the iterative optimization procedure which optimizes the Dirac mixture parameters to optimally approximate a multi-dimensional standard normal distribution with a set of equally weighted point symmetric samples.
Furthermore, we improved the numerical stability of the optimization, and together with the halved number of Dirac mixture parameters to be optimized, now it is possible to compute optimal approximations of a thousand-dimensional standard normal distribution comprising 10,000 samples.
As the Progressive Gaussian Filter (PGF) \cite{jannik_steinbring_progressive_2014} also relies on the \sskf Gaussian sampling, it can directly use and benefit from the new point symmetric sampling scheme.

The evaluations showed that the \sskf can handle symmetric measurement equations now much better when using the new symmetric sampling scheme.
It was also shown that the \sskf gave the best compromise between estimation accuracy and filter runtime when dealing with high-dimensional problems such as extended object tracking.
Additionally, this illustrated the advantage of the \sskf being able to use an arbitrary number of samples independent of the state/noise dimensions.

\appendix

\section{Odd Moments of a Point Symmetric Dirac Mixture}
\label{app:proof-odd-moments}

The odd moments of an arbitrary density function $f(\vx)$ with $\vx \in \IR^N$ are defined as
\Beq
    \E[f]{\prod_{j=1}^N x_j^{n_j}} = \int_{\IR^N} \prod_{j=1}^N x_j^{n_j} \cdot f(\vx) \dd \vx \enspace,
\Eeq
where
\Beq
    \sum_{j=1}^N n_j = 2k + 1 \,, \quad 0 \leq n_j \leq 2k + 1\,,\, k \in \IN \enspace.
\Eeq
For a standard normal distribution, i.e., $f(\vx) = \Gauss(\vx\,; \vzero, \mI_N)$, all odd moments equals zero.
Hence, we have to show that this also holds for a point symmetric Dirac mixture density function comprising $2L$ samples.
By replacing the density $f(\vx)$ with a point symmetric Dirac mixture approximation we obtain
\begin{align}
    \E[\delta]{\prod_{j=1}^N x_j^{n_j}} &= \int_{\IR^N} \prod_{j=1}^N x_j^{n_j} \frac{1}{2L} \sum_{i=1}^L \delta(\vx - \vxi ) + \delta(\vx + \vxi) \dd \vx \\
    &= \frac{1}{2L} \sum_{i=1}^L \left( \prod_{j=1}^N x_{i,j}^{n_j} + \prod_{j=1}^N (-x_{i,j})^{n_j} \right) \\[0.2cm]
    &= \frac{1}{2L} \sum_{i=1}^L \left( \prod_{j=1}^N x_{i,j}^{n_j} - \prod_{j=1}^N x_{i,j}^{n_j} \right) = 0 \enspace.
\end{align}
The same result can be easily obtained for the case of an odd number of samples $2L + 1$ where the additional sample is placed at the state space origin.

\section{Proof of Distance $D^e(S)$}
\label{app:proof-dist-even}

By using the facts that the distance $D^e(S)$ is composed of sums of products of unormalized Gaussians and their product is also an unnormalized Gaussian as well as the integral over a Gaussian equals always one, the three terms of the distance $D^e(S)$ are obtained according to
\begin{align}
    \label{eq:proof-even-d1}
    D^e_1 &= \int_0^{\bMax} \frac{1}{\pi^{\frac{N}{2}} b^{N - 1}} \int_{\IR^N} \left(\frac{b^2}{1 + b^2}\right)^{\hspace*{-0.1cm} N} \cdot (2 \pi)^N (1 + b^2)^N \Gauss(\vm\,; \vzero, (1 + b^2)\mI_N)^2 \dd \vm \; \dd b \displaybreak\\
          &= \int_0^{\bMax} \frac{1}{\pi^{\frac{N}{2}} b^{N - 1}} \left(\frac{b^2}{1 + b^2}\right)^{\hspace*{-0.1cm} N}
             \pi^\frac{N}{2} (1 + b^2)^\frac{N}{2} \dd b \\
          &= \int_0^{\bMax} b \left(\frac{b^2}{1 + b^2}\right)^{\hspace*{-0.1cm}\frac{N}{2}} \dd b \enspace,
\end{align}
\begin{align}
    \label{eq:proof-even-d2}
    D^e_2(S) =& \int_0^{\bMax} \frac{1}{\pi^\frac{N}{2} b^{N - 1}}
                \int_{\IR^N} \left(\frac{b^2}{1 + b^2}\right)^{\hspace*{-0.1cm}\frac{N}{2}}
                (2 \pi)^\frac{N}{2} (1 + b^2)^\frac{N}{2} \cdot \Gauss(\vm\,; \vzero, (1 + b^2)\mI_N) \cdot
                              \frac{(2 \pi)^\frac{N}{2} b^N}{2L} \, \cdot \\[0.2cm]
             & \sum_{i=1}^L \Gauss(\vm\,; \vsi, b^2\mI_N) +
                                           \Gauss(\vm\,; -\vsi, b^2\mI_N) \dd \vm \; \dd b \\
             =& \int_0^{\bMax} \frac{}{}
                \frac{2^N \pi^{\frac{N}{2}} b^{N + 1}}{2L (2 \pi)^\frac{N}{2}(1 + 2b^2)^\frac{N}{2}} \cdot \sum_{i=1}^L \gaussExp{\sqNorm{\vsi}}{(1 + 2b^2)} + \gaussExp{\sqNorm{-\vsi}}{(1 + 2b^2)} \dd b \\
             =& \int_0^{\bMax} \frac{2b}{2L} \left(\frac{2 b^2}{1 + 2 b^2}\right)^{\hspace*{-0.1cm}\frac{N}{2}} \cdot \sum_{i=1}^L \gaussExp{\sqNorm{\vsi}}{(1 + 2b^2)} \dd b \enspace,
\end{align}
and
\begin{align}
    \label{eq:proof-even-d3}
    D^e_3(S) &= \int_0^{\bMax} \frac{1}{\pi^\frac{N}{2} b^{N - 1}}
                \int_{\IR^N} \left( \frac{(2 \pi)^\frac{N}{2} b^N}{2L} \right)^{\hspace*{-0.1cm}2} \cdot \sum_{i=1}^L \Gauss(\vm\,; \vsi, b^2\mI_N) + \Gauss(\vm\,; -\vsi, b^2\mI_N) \,\cdot \\
             &\hspace*{6.4cm} \sum_{j=1}^L \Gauss(\vm\,; \vsj, b^2\mI_N) + \Gauss(\vm\,; -\vsj, b^2\mI_N) \dd \vm \; \dd b \\[0.2cm]
             &= \int_0^{\bMax}
                \frac{2^N \pi^\frac{N}{2} b^{N+1}}{(2L)^2 (2 \pi)^\frac{N}{2}(2b^2)^\frac{N}{2}} \sum_{i=1}^L \sum_{j=1}^L \gaussExp{\sqNorm{\vsi - \vsj}}{2b^2} + \gaussExp{\sqNorm{\vsi + \vsj}}{2b^2} + \\
             &\hspace*{6cm}\gaussExp{\sqNorm{-\vsi - \vsj}}{2b^2} + \gaussExp{\sqNorm{\vsj - \vsi}}{2b^2} \dd b \\[0.2cm]
             &= \int_0^{\bMax} \hspace*{-0.1cm} \frac{2b}{(2L)^2}
                \sum_{i=1}^L \sum_{j=1}^L \gaussExp{\sqNorm{\vsi - \vsj}}{2b^2} + \gaussExp{\sqNorm{\vsi + \vsj}}{2b^2} \dd b \enspace.
\end{align}

\section{Proof of Theorem \ref{theorem-closed-form-even-d3}}
\label{app:proof-closed-form-even-d3}

Like in \cite{uwe_d._hanebeck_dirac_2009}, to compute the term $D^e_3(S)$ we use that for $z > 0$
\Beq
    \label{eq:closed-form-integral-exp-div-b}
    \int_0^{\bMax} \frac{2}{b} \gaussExp{z}{2b^2} \dd b = -\Ei(-\frac{1}{2}\frac{z}{2\bMax^2}) \enspace,
\Eeq
where $\Ei(x)$ is the exponential integral defined as
\Beq
    \label{eq:def-expint}
    \Ei(x) = \int_{-\infty}^x \frac{e^t}{t} \dd t \enspace.
\Eeq
Moreover, the product rule gives
\Beq
    \frac{\bMax^2}{2} \gaussExp{z}{2\bMax^2} &= \int_0^{\bMax} b\, \gaussExp{z}{2b^2} \dd b + \frac{z}{4} \int_0^{\bMax}  \frac{1}{b} \gaussExp{z}{2b^2} \dd b \enspace,
\Eeq
and together with \Eq{eq:closed-form-integral-exp-div-b} we obtain
\Beq
    \label{eq:closed-form-integral-exp-times-b}
    \int_0^{\bMax} b\, \gaussExp{z}{2b^2} \dd b &= \frac{\bMax^2}{2} \gaussExp{z}{2\bMax^2} + \frac{z}{8} \Ei(-\frac{1}{2}\frac{z}{2\bMax^2}) \enspace.
\Eeq
This directly results in the expression
\Beq
    \label{eq:proof-closed-form-even-d3}
    D^e_3(S) &= \frac{2}{(2L)^2} \sum_{i=1}^L \sum_{j=1}^L
        \frac{\bMax^2}{2}
        \left( \gaussExp{\sqNorm{\vsi - \vsj}}{2\bMax^2} + \gaussExp{\sqNorm{\vsi + \vsj}}{2\bMax^2} \right) \,+ \\
        &\hspace*{2.8cm} \frac{1}{8} \left(\sqNorm{\vsi - \vsj} \Ei\left(-\frac{1}{2}\frac{\sqNorm{\vsi - \vsj}}{2\bMax^2}\right) + \sqNorm{\vsi + \vsj} \Ei\left(-\frac{1}{2}\frac{\sqNorm{\vsi + \vsj}}{2\bMax^2}\right)\right) \enspace.
\Eeq

\section{Proof of Distance $D^o(S)$}
\label{app:proof-dist-odd}

The distance $D^o(S)$ differs from its even counterpart due to the additional sample placed fixed at the state space origin.
This does not effect $D^o_1$, and hence, it equals $D^e_1$.
The other two terms are sums of their reweighted even counterparts (due to the changed sample weight) and terms comprising also products of unnormalized Gaussians.
Hence, they are given as
\begin{align}
    \label{eq:proof-odd-d2}
    D^o_2(S) &= \frac{2L}{2L + 1} D^e_2(S) + \int_0^{\bMax} \frac{1}{\pi^\frac{N}{2} b^{N - 1}}
                \int_{\IR^N} \left(\frac{b^2}{1 + b^2}\right)^{\hspace*{-0.1cm}\frac{N}{2}} \cdot \\
             &\hspace*{0.4cm} (2 \pi)^\frac{N}{2} (1 + b^2)^\frac{N}{2}
                \Gauss(\vm\,; \vzero, (1 + b^2)\mI_N) \,\cdot \frac{(2 \pi)^\frac{N}{2} b^N}{2L + 1}
                \Gauss(\vm\,; \vzero, b^2\mI_N) \dd \vm \; \dd b \\
             &= \frac{2L}{2L + 1} D^e_2(S) + \int_0^{\bMax} \frac{2^N \pi^{\frac{N}{2}} b^{N + 1}}{2L + 1}
                \frac{1}{(2 \pi)^\frac{N}{2}(1 + 2b^2)^\frac{N}{2}} \dd b \\[-0.05cm]
             &= \frac{2L}{2L + 1} D^e_2(S) + \int_0^{\bMax} \hspace*{-0.1cm} \frac{b}{2L + 1}
                \left(\frac{2 b^2}{1 + 2 b^2}\right)^{\hspace*{-0.1cm}\frac{N}{2}} \dd b
\end{align}
and
\begin{align}
    \label{eq:proof-odd-d3}
    D^o_3(S) &= \frac{(2L)^2}{(2L + 1)^2} D^e_3(S) +
                \int_0^{\bMax} \frac{1}{\pi^\frac{N}{2} b^{N - 1}} \cdot \int_{\IR^N} \left( \frac{(2 \pi)^\frac{N}{2} b^N}{2L + 1} \right)^{\hspace*{-0.1cm}2} \,\cdot \\
             &\hspace*{0.4cm} \Bigg( 2 \cdot \Gauss(\vm\,; \vzero, b^2\mI_N) \cdot \left(\sum_{i=1}^L \Gauss(\vm\,; \vsi, b^2\mI_N) +
                \Gauss(\vm\,; -\vsi, b^2\mI_N)\right) + \Gauss(\vm\,; \vzero, b^2\mI_N)^2 \Bigg) \dd \vm \; \dd b \\
             &= \frac{(2L)^2}{(2L + 1)^2} D^e_3(S) + \int_0^{\bMax}
                \frac{2^N \pi^\frac{N}{2} b^{N+1}}{(2L + 1)^2} \,\cdot \\
             &\hspace*{0.4cm} \frac{1}{(2 \pi)^\frac{N}{2}(2b^2)^\frac{N}{2}}
                \Bigg(2 \cdot \sum_{i=1}^L \gaussExp{\sqNorm{\vsi}}{2b^2} + \gaussExp{\sqNorm{-\vsi}}{2b^2} + 1\Bigg) \dd b \\
             &= \frac{(2L)^2}{(2L + 1)^2} D^e_3(S) +
                \frac{\bMax^2}{2 (2L + 1)^2} + \int_0^{\bMax} \frac{4b}{(2L + 1)^2}
                \sum_{i=1}^L \gaussExp{\sqNorm{\vsi}}{2b^2} \dd b \enspace.
\end{align}

\section{Proof of Theorem \ref{theorem-closed-form-odd-d3}}
\label{app:proof-closed-form-odd-d3}

A closed-form expression for $D^o_3(S)$ can directly be obtained by using again \Eq{eq:closed-form-integral-exp-times-b} as well as the closed-form expression for $D^e_3(S)$ resulting in
\Beq
    \label{eq:proof-closed-form-odd-d3}
    D^o_3(S) &= \frac{(2L)^2}{(2L + 1)^2} D^e_3(S) + \frac{\bMax^2}{2 (2L + 1)^2} \,+ \\
             &\hspace*{0.4cm} \frac{4}{(2L + 1)^2} \sum_{i=1}^L \frac{\bMax^2}{2} \gaussExp{\sqNorm{\vsi}}{2\bMax^2} + \frac{1}{8} \sqNorm{\vsi} \Ei\left(-\frac{1}{2}\frac{\sqNorm{\vsi}}{2\bMax^2}\right) \enspace.
\Eeq

\section{Boundedness of $D^e(S)$ and $D^o(S)$}
\label{app:proof-boundedness}

We show the boundedness of the distances $D^e(S)$ and $D^o(S)$ for an increasing dimension $N$.
For a given $\bMax$ it holds
\Beq
    \lim_{N \to \infty} D^e_1 &= \lim_{N \to \infty} \int_0^{\bMax} b \underbrace{\left(\frac{b^2}{1 + b^2}\right)^{\hspace*{-0.1cm}\frac{N}{2}}}_{\to 0 \;\text{for } N \to \infty} \dd b = 0 \enspace,
\Eeq
\Beq
    \lim_{N \to \infty} D^e_2(S) &= \lim_{N \to \infty} \int_0^{\bMax} \frac{2b}{2L} \left(\frac{2 b^2}{1 + 2 b^2}\right)^{\hspace*{-0.1cm}\frac{N}{2}} \cdot \underbrace{\sum_{i=1}^L \gaussExp{\sqNorm{\vsi}}{(1 + 2b^2)}}_{\leq L} \dd b \\
        &\leq \lim_{N \to \infty} \int_0^{\bMax} b \underbrace{\left(\frac{2 b^2}{1 + 2 b^2}\right)^{\hspace*{-0.1cm}\frac{N}{2}}}_{\to 0 \;\text{for } N \to \infty} \dd b = 0 \enspace,
\Eeq
and
\begin{align}
    \lim_{N \to \infty} D^e_3(S) &= \lim_{N \to \infty} \int_0^{\bMax} \hspace*{-0.1cm} \frac{2b}{(2L)^2} \cdot \underbrace{\sum_{i=1}^L \sum_{j=1}^L \gaussExp{\sqNorm{\vsi - \vsj}}{2b^2} + \gaussExp{\sqNorm{\vsi + \vsj}}{2b^2}}_{\leq 2 L^2} \dd b \\
        &\leq \lim_{N \to \infty} \int_0^{\bMax} b \dd b = \frac{\bMax^2}{2} \enspace.
\end{align}
Hence, the distance $D^e(S)$ is bounded by $\bMax$ according to
\Beq
    \lim_{N \to \infty} D^e(S) = \lim_{N \to \infty} D^e_1 -2 D^e_2(S) + D^e_3(S) \leq \frac{\bMax^2}{2} \enspace.
\Eeq
In a similar manner, the same result can be obtained for the distance $D^o(S)$.

\section{Proof of Theorem \ref{theorem-closed-form-even-deriv-d3}}
\label{app:proof-closed-form-even-deriv-d3}

A closed-form expression for $\frac{\partial D^e_3(S)}{\partial \sid}$ can directly be obtained using \Eq{eq:closed-form-integral-exp-div-b} resulting in
\Beq
    \label{eq:proof-closed-form-even-deriv-d3}
    \frac{\partial D^e_3(S)}{\partial \sid} &= \frac{1}{(2L)^2} \sum_{j=1}^L (\sid - \sjd) \Ei\left(-\frac{1}{2}\frac{\sqNorm{\vsi - \vsj}}{2\bMax^2}\right) + (\sid + \sjd) \Ei\left(-\frac{1}{2}\frac{\sqNorm{\vsi + \vsj}}{2\bMax^2}\right) \enspace.
\Eeq

\section{Proof of Theorem \ref{theorem-closed-form-odd-deriv-d3}}
\label{app:proof-closed-form-odd-deriv-d3}

A closed-form expression for $\frac{\partial D^o_3(S)}{\partial \sid}$ can directly be obtained using~\Eq{eq:closed-form-integral-exp-div-b} and the closed-form expression for $\frac{\partial D^e_3(S)}{\partial \sid}$ resulting in
\Beq
    \label{eq:proof-closed-form-odd-deriv-d3}
    \frac{\partial D^o_3(S)}{\partial \sid} &= \frac{(2L)^2}{(2L + 1)^2} \frac{\partial D^e_3(S)}{\partial \sid} + \frac{\sid}{(2L + 1)^2} \Ei\left(-\frac{1}{2}\frac{\sqNorm{\vsi}}{2\bMax^2}\right)  \enspace.
\Eeq

    \bibliographystyle{IEEEtran}
    \bibliography{Literature}
\end{document}